\RequirePackage{fix-cm}
\documentclass[smallextended]{svjour3}       
\smartqed  
\usepackage{graphicx}
\usepackage{natbib}
\usepackage{amsmath,amssymb}

\newcommand{\Ks}{{K_{\rm s}}}
\newcommand{\JHKs}{{JH\Ks}}
\newcommand{\EHKs}{{E_{H-\Ks}}}
\newcommand{\AKs}{{A_{\Ks}}}
\newcommand{\AKEHK}{{\AKs}/\EHKs}
\newcommand{\RGC}{{R_{\rm GC}}}
\newcommand{\Teff}{{T_{\rm eff}}}
\newcommand{\aj}{AJ}

\newcommand{\araa}{ARA\&A}
\newcommand{\apj}{ApJ}
\newcommand{\apjl}{ApJ}
\newcommand{\apjs}{ApJS}

\newcommand{\apss}{Ap\&SS}
\newcommand{\aap}{A\&A}

\newcommand{\aaps}{A\&AS}

\newcommand{\memras}{MmRAS}
\newcommand{\mnras}{MNRAS}
\newcommand{\na}{New A}
\newcommand{\nar}{New A Rev.}

\newcommand{\pasp}{PASP}
\newcommand{\pasj}{PASJ}

\newcommand{\rmxaa}{Rev. Mexicana Astron. Astrofis.}

\newcommand{\ssr}{Space~Sci.~Rev.}
\newcommand{\zap}{ZAp}
\newcommand{\nat}{Nature}

\newcommand{\bain}{Bull.~Astron.~Inst.~Netherlands}

\newcommand{\procspie}{Proc.~SPIE}
\begin{document}

\title{Impact of distance determinations on Galactic structure.~I. Young and intermediate-age tracers
}


\author{
Noriyuki Matsunaga$^{1}$ \and Giuseppe Bono$^{2,3}$ \and Xiaodian Chen$^{4}$ \and Richard de Grijs$^{5,6,7}$ \and Laura Inno$^{8,2}$ \and Shogo Nishiyama$^{9}$ 
}

\authorrunning{N. Matsunaga et al.} 

\institute{
N. Matsunaga \at
  \email{matsunaga@astron.s.u-tokyo.ac.jp}           
\and
Giuseppe Bono \at
  \email{bono@roma2.infn.it}           
\and
Xiaodian Chen \at
  \email{chenxiaodian@nao.cas.cn}           
\and
Richard de Grijs \at
  \email{richard.de-grijs@mq.edu.au}           
\and
L. Inno \at
  \email{inno@mpia.de}           
\and
Shogo Nishiyama \at
  \email{shogo-n@staff.miyakyo-u.ac.jp}           
\and
$^{1}$ Department of Astronomy, The University of Tokyo, 7-3-1 Hongo, Bunkyo-ku, Tokyo 113-0033, Japan  \\
$^{2}$ Dipartimento di Fisica, Universit\'a di Roma Tor Vergata, Via della Ricerca Scientifica 1, I-00133 Rome, Italy \\
$^{3}$ INAF--Oservatorio Astronomico di Roma, Via Frascati 33, 00078 Monte Porzio Catone, Italy \\
$^{4}$ Key Laboratory for Optical Astronomy, National Astronomical Observatories, Chinese Academy of Sciences, 20A Datun Road, Chaoyang District, Beijing 100012, China \\
$^{5}$ Kavli Institute for Astronomy \& Astrophysics and Department of Astronomy, Peking University, Yi He Yuan Lu 5, Hai Dian District, Beijing 100871, China \\
$^{6}$ Department of Physics and Astronomy, Macquarie University, Balaclava Road, North Ryde, NSW 2109, Australia  \\
$^{7}$ International Space Science Institute, Beijing, 1 Nanertiao, Zhongguancun, Hai Dian District, Beijing 100190, China \\
$^{8}$ Max-Planck Institute for Astronomy, K\"onigstuhl 17, D-69117, Heidelberg, Germany \\
$^{9}$ Miyagi University of Education, Aoba-ku, Sendai, Miyagi 980-0845, Japan
}

\date{Received: date / Accepted: date}

\maketitle

\begin{abstract}
Here we discuss impacts of distance determinations on
the Galactic disk traced by relatively young objects.
The Galactic disk, {$\sim$}40~kpc in diameter, is a cross-road of studies on
the methods of measuring distances,
interstellar extinction, evolution of galaxies,
and other subjects of interest in astronomy.
A proper treatment of interstellar extinction is, for example,
crucial for estimating distances to stars
in the disk outside the small range of the solar neighborhood.
We'll review the current status of relevant studies
and discuss
some new approaches to the extinction law.
When the extinction law is reasonably constrained,
distance indicators found in today and future surveys are 
telling us stellar distribution and more throughout the Galactic disk.
Among several useful distance indicators, the focus of this review is
Cepheids and open clusters (especially contact binaries in clusters).
These tracers are particularly useful for addressing the metallicity
gradient of the Galactic disk,
an important feature for which comparison between
observations and theoretical models
can reveal the evolutionary of the disk.

\keywords{Stars: variables \and Cepheids: distance scale}
\end{abstract}

\section{Introduction}
\label{sec:Intro}

Understanding the Galaxy is important not only because it hosts
our Solar system but also because we can investigate stellar populations
and interstellar components in greater detail than in any other galaxies, and
the Galaxy can be a benchmark for galaxy evolution. Recent large surveys
of stellar populations in various regions of the Galaxy have been
revolutionizing the study of its structure and evolution
\citep[e.g.][]{Ivezic-2012,Bland-Hawthorn-2016}.
In addition to many exciting results on the halo, large-scale
photometric and spectroscopic surveys also make it possible to
test chemodynamical models of the disk with various physical processes like 
radial migration \citep[e.g.][]{Schonrich-2009,Casagrande-2011,Haywood-2013}. 
Radial migration is the mixing process induced 
by the non-axisymmetric potential of the Galaxy,
which brings outwards the metal-rich objects 
formed in the inner disk and vice-versa \citep{Sellwood-2002,Grand-2015}.
We need stellar tracers whose distances can be well measured over
the large volume of the disk to study such a large-scale phenomenon.
In this context, classical Cepheids and open clusters
are useful distance indicators
in order to trace the young stellar components of the disk.
In particular, classical Cepheids and contact binaries found in open clusters
are useful as we will see below.
These objects are predominantly found in the disk.

\subsection{Classical Cepheids\label{sec:IntroCep}}

In this review, we focus on classical Cepheids, not type II Cepheids
which are similar pulsating stars but belong to old stellar populations
\citep{Sandage-2006,Beaton-2018}.
Classical Cepheids (hereinafter simply called Cepheids)
are pulsating stars of 4--10~M$_{\odot}$, 
which show well-defined relations between their intrinsic luminosity and the pulsation period.
Such period--luminosity (PL) relations make them fundamental calibrators of 
the extragalactic distance scale
\citep[see the review by][]{Subramanian-2017}.
Besides this important role, Cepheids also have a number of properties
that make them ideal probes of both the structure 
and the recent history of the Galaxy.
In fact, Cepheids are luminous stars that can be seen to great distances even
through the substantial interstellar extinction along the lines of sight.
In addition to the distances, the ages of individual Cepheids can be
determined on the basis of their periods \citep{Bono-2005}. They are
young stars (10--200~Myr), comparable with the characteristic dynamical
time-scale or the orbital period ({$\sim$}240~Myr at the solar location
if one assumes the IAU recommended values of
the Sun--Galactic Center distance $R_0=8.5$~kpc and
the rotation speed $V_0=220$~km~s$^{-1}$)
of the Galactic disk, thus they are still relatively close
to the place where they were born as far as the distances from
the Galactic Center 
(henceforth the GC) and to neighboring objects are concerned.
These make the Cepheids ideal tracers for studying
dynamical evolution of the young components like spiral arms
\cite[see e.g.][]{Baba-2018,Kawata-2018}.
Furthermore, they are good chemical tracers.
Their effective temperatures ($\Teff \sim 5500$~K) 
are relatively low when compared to similarly young stars,
and allow us to measure many metallic lines in their spectra
giving access to precise abundances of many different elements
\citep[e.g.][]{Lemasle-2013,daSilva-2016}. 

\subsection{Open clusters}
\label{sec:IntroOC}

Open clusters are aggregates of hundreds to tens of thousands of
stars, all born together. Their members have similar ages,
metallicities, and distances to us, which means that they usually show
obvious main sequences in the Hertzsprung--Russell diagram. Based on
the main sequence of a cluster, a unique theoretical isochrone can be
identified to describe the evolutionary details of stars with
different masses. In contrast to the Galactic globular clusters
which are predominantly old, the ages of open clusters range from
very young (a few Myr) to rather old (a few Gyr), and they are
mostly located in the disk. Since their distances can be 
determined based on main-sequence or isochrone fitting,
open clusters represent important tools to trace young structures such as
spiral arms \citep[see, e.g.][]{Dias-2005,Carraro-2014}.
Combining the proper motions, radial
velocities, and metallicity information, open clusters can be used to study the
kinematics and evolution of the young components of the Galaxy.

Furthermore, studies of variable stars in open clusters are interesting.
In addition to  
Cepheids and other pulsating stars, eclipsing contact binaries are
another type of variable often found in open clusters.
One can expect
PL (or PLC) relations for contact binaries (see \S\ref{sec:CBOC}),
and recent studies actually found relatively tight relations,
especially in the near-IR \citep{Chen-2016b}.
Such distance indicators, combined with the isochrone fitting method,
can provide mutual verification of distance scales.

\subsection{Other young tracers}

There are a few other young and intermediate-age tracers
which are useful to study Galactic structure.
\begin{itemize}
\item {\it Massive star-forming regions} often host
methanol (CH$_3$OH) masers which enable us
to measure the trigonometric parallaxes of the star-forming regions
with VLBI (Very Long Baseline Interferometry) facilities
\citep{Reid-2009,Honma-2012,Reid-2014a}.
The high accuracy of VLBI measurements, in the best cases
better than 10~{$\mu$as} \citep[e.g.][]{Zhang-2013,Sanna-2017}, allows us to map
this youngest population over a large volume of the disk.

\item {\it Long-period variables} (LPVs hereinafter)
include evolved asymptotic giant branch (AGB) stars with
a wide range of ages between 100~Myr and over 10~Gyr.
The lower limit to their ages, or the upper limit to
the initial mass with which a star evolves into an AGB star,
instead of a red supergiant, is unclear.
LPVs (or also called Miras) obey PL relations
and thus serve as distance indicators
\citep{Whitelock-2008,Subramanian-2017}.
Their high luminosities, especially in the infrared (hereinafter IR),
make them useful tracers of the Galaxy's structure. 
Some of them also show radio maser emission lines which are
useful to study the kinematics of the Galaxy.
In particular, a group of relatively massive LPVs
(3--5~$M_\odot$, age $< 1$~Gyr) are often called OH/IR stars
because they have OH masers and are very bright in the IR.
They have provided us with good insights, for example, into
the inner part of the Galactic disk \citep{Lindqvist-1992,Sevenster-2000}. 

\item {\it Red clump (RC) giants} are also useful tracers of
stellar populations and very widely used
for studying Galactic structure.
They are core-helium-burning stars found spanning a wide range of 
ages from hundreds of Myr to 10~Gyr. Note that they are found as
a clump in the color-magnitude diagram or in a luminosity function
so that they can be identified only if a sizeable number of such stars
in stellar population(s) are present.
The specific frequency of the core-helium-burning stars,
the number per unit stellar mass of 
the population, is maximum around 1~Gyr \citep{Salaris-2002}.
Taking the total mass of stars at different ages into account, however,
the age distribution of the RC stars tends to be extended
unless the star formation history is represented by a sharp peak. 
The brightness of the clump depends on the age and metallicity, 
but the dependency is mild against 
the rather wide range of their ages, 1.5--10~Gyr \citep{Salaris-2002}.
The error of distance estimation caused by the age distribution of stars
tends to be averaged out,
which is an advantage in estimating distances to systems
with various stellar populations mixed,
but makes it hard to separate RCs formed by stars with different ages.
These complicate the use of the RC for 
studying the distribution of stars at a specific age in the Galaxy.

\end{itemize}

These objects are not discussed in the rest of the review. 
On massive star-forming regions, readers are referred to
\citet[][and references therein]{Reid-2014b}.
On LPVs and RC giants, \citet{Subramanian-2017} discuss
their characteristics and applications.
Some other kinds of young stars are also investigated as
distance indicators, e.g.\  blue supergiants
\citep{Kudritzki-2003,Kudritzki-2008} and
red supergiants \citep{Yang-2012}.
In the following we focus on Cepheids and open clusters. 
We can measure the individual distances
to these objects based on optical to IR photometric data,
which is important for characterizing the interstellar extinction
that shows very patchy patterns on the sky.

\subsection{The goal of this review}

The purpose of this review is to discuss the impact of
distance determinations of Cepheids and open clusters
(Cepheids and contact binaries in the clusters, in particular) for 
revealing the structure and evolution of the disk,
which is a relatively unexplored frontier of Galactic archeology
mainly because many previous surveys in the optical wavelengths
were seriously affected by interstellar extinction.

After their birth, stars in the Galaxy change their orbits through
secular evolution processes, but their chemical abundances remain
almost unaltered from their compositions at birth 
(except for a few elements that may be affected by stellar evolution processes;
for Cepheids, e.g., see \citealt{Takeda-2013}).
This motivates the science of Galactic archeology, 
a recently emerging field in astronomy which aims at
studying the imprints of chemodynamical 
evolution from the chemistry and kinematics of a large number of stars.
In fact, we are now experiencing a quantum jump both
in the wealth and in the quality of available information
for billions of stars in the Galaxy
thanks to the efforts of dedicated spectroscopic surveys
(e.g~APOGEE, GALAH, LAMOST, Gaia-ESO, and in the future 4MOST),
capable of obtaining
detailed chemical compositions, and to some extent distances and ages,
of individual evolved stars (e.g.\  RC giants, {$\sim$}1--10 Gyr,
and red giant branch stars, {$\sim$}10--13~Gyr) throughout
the entire Galactic disk.  

However, a similarly detailed map of the
interstellar gas and young stars is still lacking.
While we can obtain abundances of only a few elements (e.g.~oxygen and nitrogen)
for interstellar gas, we can measure the detailed abundance for 
the young stars,
which should still reflect the composition of their birth material.
A critical observable feature for investigating the large-scale disk
evolution is the metallicity gradient,
i.e.~metallicity distribution
as a function of Galactocentric distance ($\RGC$). 
In fact, an inside-out formation scenario for the Galactic disk
will produce metal-rich stars in the inner part
and metal-poor ones in the outer part of the Galactic disk
\citep[e.g.][and references therein]{Tsujimoto-1995,Chiappini-2001}.
Such a trend observed in interstellar gas and youngest stars
reflects the present-epoch and recent chemical structure of the disk,
and the time-dependent trend traced in stars with various ages can tell us
the chemical structure in the past and the evolution of the disk.
However, observational reconstruction of the metallicity gradient
as a function of time is challenging, at least,
for the following two reasons: (1)~Good tracers with 
accurate estimates of distance, age, and metallicity need to be found and
well characterized. (2)~The interstellar extinction hampers accurate
distance determination and detailed observations of tracers spread
across the disk. Young tracers like Cepheids and open clusters 
embedded in the Galactic disk, and
those at large distances, several kpc or more distant,
are severely affected by extinction as we will see below.

In this review, we will summarize up-to-date observational results
and current problems about the interstellar extinction in \S\ref{sec:Ext}.
Then, recent developments of the distance measurements and some important problems
are discussed for Cepheids in the Galactic disk ({\S}\ref{sec:CepDisk}),
Cepheids in open clusters ({\S}\ref{sec:CepOC}),
and contact binaries in open clusters (\S\ref{sec:CBOC}).
Then, in \S\ref{sec:gradient}, we discuss how the distance determination
and the characterization of the interstellar extinction are important to study 
the metallicity gradient. Section~\ref{sec:Remarks} summarizes the review.

\section{The Interstellar Extinction Law}
\label{sec:Ext}

A detailed understanding of the interstellar extinction law,
or the wavelength dependence of the interstellar extinction, 
is crucial to most aspects of observational astronomy.
Dust grains attenuate the light from an object seen through them,
and the amount of attenuation depends on wavelength
and the lines of sight.
Equations that fit the observed wavelength dependence of the extinction 
are required to test dust grain models
and to predict the amount of extinction in certain spectral regions.

A large impact of the extinction law on Galactic structure
can be highlighted by comparing the {\it classical} law
in \citet{Cardelli-1989}, for example, with those derived in recent studies.
\citet{Nishiyama-2008} reported an extinction ratio of
$A_V / A_{K_{\rm s}} \approx 16$, a factor of {$\sim$}1.9 larger than
the value of $8.5$ in \citet{Cardelli-1989} for $R_V \equiv A_V/E_{B-V} = 3.1$.
Considering the column density toward the GC obtained in X-ray observations
\citep{Porquet-2008}, \citet{Fritz-2011} derived a lower limit of
$A_V / A_{K_{\rm s}} \approx 15$. The difference between the new results and
previous ones \citep{Rieke-1985,Cardelli-1989} can be clearly
seen in Fig.~8 in \citet{Fritz-2011}.
In addition,  a recent study by \citet{Hosek-2018} claims that even
\citet{Nishiyama-2008} underestimated the optical-to-near IR extinction ratio.
These results suggest that when we use the {\it classical} extinction law
where it is not appropriate,
the amount of extinction may be associated with an error of more than 50~\%.

In this section, we review the extinction law 
in wavelength order:
the ultra-violet (UV) and optical range (\S\ref{sec:ExtUVOpt}),
the near IR  (\S\ref{sec:ExtNIR}),
and the mid IR  (\S\ref{sec:ExtMIR}).
In the UV and optical wavelengths, 
the difference of the interstellar extinction law among various lines
of sight was already known in early studies
\citep[e.g.][]{Johnson-1955,Johnson-1963,Nandy-1965}.
For several decades, in contrast,
the extinction law in the near- and mid-IR wavelengths 
was believed to show a very little, if any, variation from
one line of sight to another.
The ``universality'' of the IR extinction law implies that
the dust grains responsible for the near- and mid-IR extinction
have almost the same size distributions along all sightlines.
In addition, the IR extinction was thought to
follow a power law of $A_{\lambda} \propto \lambda^{-\alpha}$
from 1~{$\mu$m} to $\sim 8$~{$\mu$m} in early studies
\citep[e.g.][]{Draine-1989}.
Some recent results, however, suggest that the IR extinction law 
is not so simple as we will see in this section,
while there are still results supporting the ``universality''
of the IR extinction law (\S\ref{sec:Remarks}).

\subsection{The interstellar extinction law: UV and optical}
\label{sec:ExtUVOpt}

\citet{Johnson-1963} found a variation of 
the ratio of the total-to-selective extinction in the $B$ and $V$ bands,
$R_V$
from 3.1 to 7.4. 
\citet{Bless-1972} examined the extinction law from the UV to the near-IR
for bright early-type stars,
and clearly demonstrated large variations of the extinction law.
The parameter $R_V$ 
depends on the dust properties along the lines of sight.
Large grains produce gray extinction, i.e., small wavelength dependence,
leading to large $R_V$.

In spite of such large variations of the UV/optical extinction law,
\citet{Cardelli-1989} suggested that 
the variation of the extinction law at UV--optical wavelengths
can be described by a functional form with only one parameter, $R_V$.
Their suggestion was based on the fact that
there is a good linear correlation between
$A_{\lambda}/A_V$ and $R_V^{-1}$
($A_{\lambda}/A_V = a_{\lambda} + b_{\lambda}R_V^{-1}$)
over an entire range of wavelengths from the UV to the near-IR.
\citet{Cardelli-1989} provided the $\lambda$-dependent coefficients, 
$a_{\lambda}$ and $b_{\lambda}$, from the UV to the near-IR, 
which allows us to calculate $A_{\lambda}/A_V$ 
if the value of $R_V$ is known for the sight line.

The one-parameter extinction law was welcomed by the astronomical community.
This makes it easier to correct the amount of extinction
not only in the commonly used photometric bands actually measured
by \citet{Cardelli-1989} but for any band or wavelength.
However, \citet{Fitzpatrick-2007} pointed out that 
the correlation found by \citet{Cardelli-1989} is ``partially illusory'',
and the extinction law is not a one-parameter function.
The functional form of \citet{Cardelli-1989} was proposed
because of a good correlation between $A_{\lambda}/A_V$ and $R_V^{-1}$.
However, $A_{\lambda}/A_V$ was determined based on 
the color-excess ratio $E_{\lambda -V}/E_{B-V}$ and $R_V$ as
\[
\frac{A_{\lambda}}{A_V} = \frac{E_{\lambda -V}}{E_{B-V}} R_V^{-1} + 1.
\]
Since $R_V$ is used to calculate $A_{\lambda}/A_V$,
some degree of apparent correlation is expected between
$A_{\lambda}/A_V$ and $R_V^{-1}$ in any case.

More than 300 curves for the extinction law
from UV to optical wavelengths were obtained
for different objects by \citet{Fitzpatrick-2007}.
The curves range from nearly gray to 
strongly $\lambda$-dependent.
\citet{Fitzpatrick-2007} concluded that 
there is no functional relationship between $R_V$ and the UV extinction law,
and that the UV extinction cannot be represented by only a single parameter.
It is possible to derive a typical or average extinction law;
however, any mean curve is subject to 
a bias caused by the selection of sample
used for producing the curve.
Hence their conclusion is that there is {\it no} unique or best estimate
of the mean extinction law in our Galaxy, at least, at UV--optical wavelengths.

\subsection{The interstellar extinction law: near-IR}
\label{sec:ExtNIR}

\begin{figure}[h]
  \includegraphics[bb=0 0 451 311,width=0.65\textwidth]{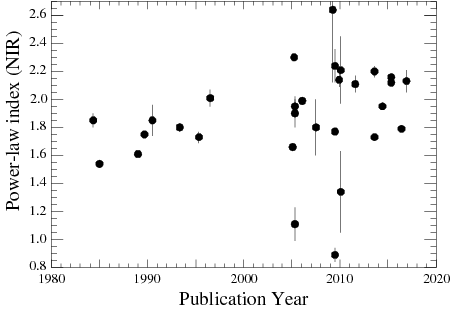}
\caption{
  The power-law index $\alpha$ for the near-IR extinction law
  is plotted against publication year since 1984.
  The largest and smallest $\alpha$ values are both from \citet{Moore-2005}.
  References:
  \citet{Landini-1984},
  \citet{Rieke-1985},
  \citet{Cardelli-1989},
  \citet{Draine-1989},
  \citet{Martin-1990},
  \citet{Whittet-1993},
  \citet{He-1995},
  \citet{Lumsden-1996},
  \citet{Larson-2005},
  \citet{Moore-2005},
  \citet{Indebetouw-2005},
  \citet{Messineo-2005},
  \citet{Nishiyama-2006},
  \citet{Froebrich-2007},
  \citet{Gosling-2009},
  \citet{Fitzpatrick-2009},
  \citet{Stead-2009},
  \citet{Schodel-2010},
  \citet{Fritz-2011},
  \citet{Wang-2013},
  \citet{Wang-2014a},
  \citet{MaizApellaniz-2015},
  \citet{Schultheis-2015},
  \citet{Xue-2016},
  and \citet{Damineli-2016}.
}
\label{fig:NIRAlpha}
\end{figure}

It has been considered that the near-IR extinction law can be approximated
by the power-law form, $A_{\lambda} \propto \lambda^{-\alpha}$.
Fig.~\ref{fig:NIRAlpha} shows the power-law index $\alpha$
as a function of publication year since 1984.
Before $\sim 1995$, many studies had converged to the IR extinction law
that is nearly constant and universal (with $\alpha \approx 1.6 - 1.8$).
However, steeper extinction law has been derived in many recent works,
e.g.~$\alpha = 2.3$ for stars behind translucent clouds
at high Galactic latitude \citep{Larson-2005},
1.9 for late-type stars with SiO maser emission \citep{Messineo-2005},
1.99 toward the GC \citep{Nishiyama-2006}.

A simple comparison of the extinction law in different lines of sight
can be done by using reddening vectors in color-color diagrams.
To determine slopes of reddening vectors precisely, 
\citet{Kenyon-1998} introduced a reddening probability function,
in which the distribution of strongly reddened stars in a color-color diagram
is compared with those of stars in a reference (off-cloud) field.
The derived slopes are $E_{J-H}/E_{H-K} = 1.57 \pm 0.03$ 
for the $\rho$ Ophiuchi dark cloud \citep{Kenyon-1998},
$1.80 \pm 0.03$ for the Chameleon~I cloud \citep{Gomez-2001},
and $2.08 \pm 0.03$ for Coalsack Globule 2 \citep{Racca-2002}.
Using the same technique but different telescopes and instruments, 
\citet{Naoi-2006,Naoi-2007} also obtained a steeper reddening slope in  
$E_{J-H}/E_{H-K_{\rm s}}$ toward Coalsack Globule 2 than in other dark clouds.
In addition, they found a change in the reddening slope;
smaller slopes are seen toward sightlines with larger optical depth. 
This trend had been implied by previous studies \citep[e.g.][]{Kenyon-1998}.
These results suggest spatial variation of the near-IR extinction law,
at least, from a line of sight to another.

\begin{figure}
  \includegraphics[width=0.55\textwidth]{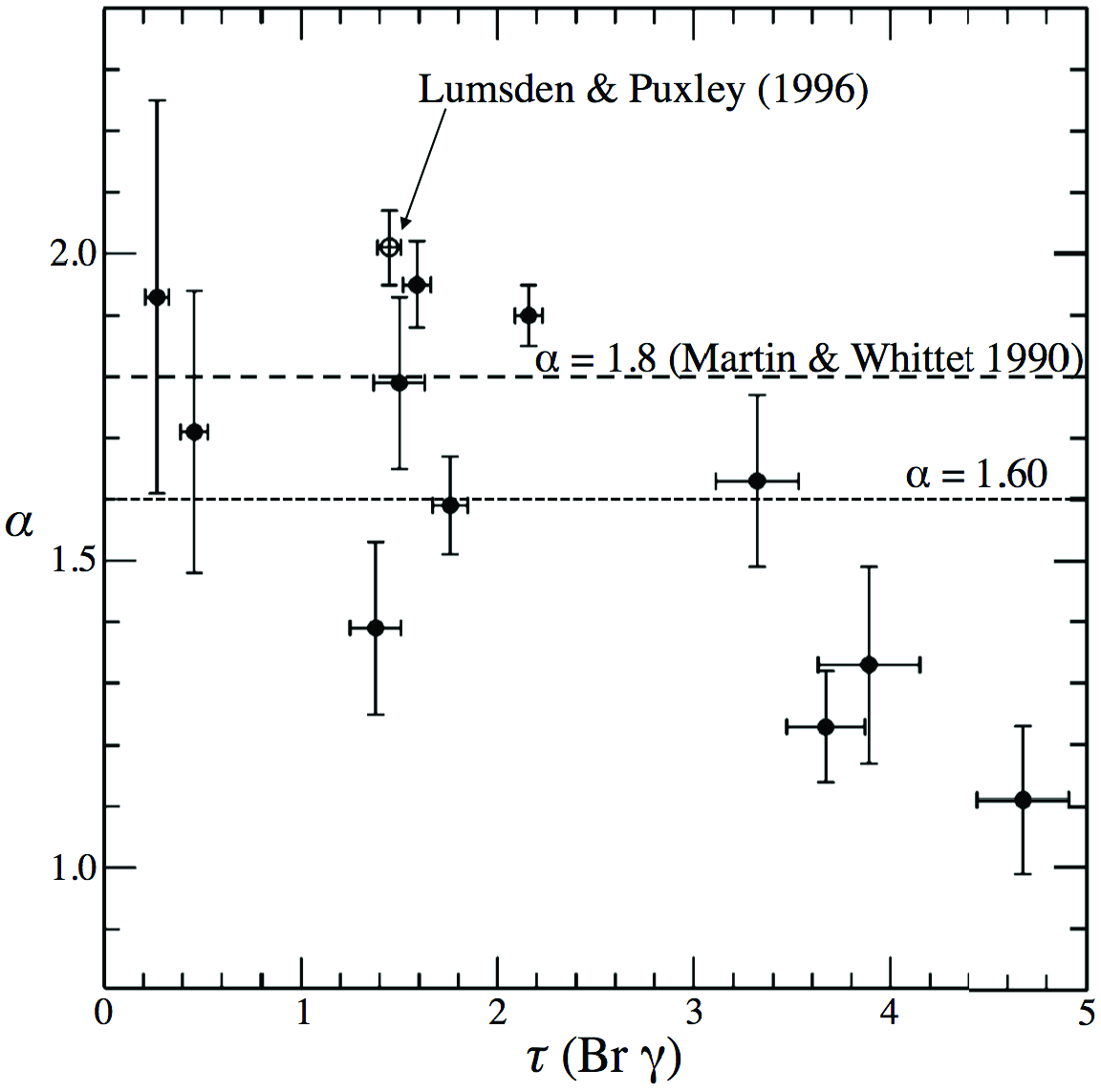}
\caption{
  The power-law index $\alpha$ versus $\tau$(Br$\gamma$).
  The indices are derived by using hydrogen lines
  for nine H\,{\small I\hspace{-.1em}I} regions and two planetary nebulae.
  A clear variation of $\alpha$ is seen.
  Adapted from \citet[][Fig.~4]{Moore-2005}.
}
\label{fig:05MooreFig4}
\end{figure}

Spectroscopic studies of the near-IR extinction law
for nine ultra compact H~II regions and two planetary nebulae
were carried out by \citet{Moore-2005}.
The amounts of extinction for their target regions are larger than $A_V=15$~mag.
They derived the power-law index $\alpha$ by using hydrogen line ratios,
and found that $\alpha$ varies from 1.11 to 1.95, with a mean of 1.6,
depending on $\tau$(Br$\gamma$) as seen in (Fig.~\ref{fig:05MooreFig4}). 

\citet{Fitzpatrick-2009} examined extinction law 
from UV to near-IR wavelengths for 14 early-type stars.
They fit the extinction law in the form (their Equation 2)
\[
\frac{E_{\lambda -V}}{E_{B-V}} = k_{\mathrm{IR}} \lambda^{-\alpha} - R_V.
\]
When the three parameters, $k_{\mathrm{IR}}$, $\alpha$, and $R_V$
are left free in the fits, $\alpha$ varies from 0.89 to 2.24
(with a mean value of 1.77 and a typical uncertainty of 0.15).
According to their conclusion, the extinction law from the $I$ band
to the $K$ band is well represented by a simple power-law for each sightline,
but $\alpha$ varies significantly among 14 lines of sight,
and there is no universal power-law index for the near-IR extinction law.
They also found that $\alpha$ depends on $R_V$,
$\alpha$ being smaller for a larger $R_V$.
This trend is consistent with the $\alpha$ dependency on $\tau$(Br$\gamma$)  
found by \citet{Moore-2005} because larger $R_V$ values are
found toward lines of sight with strong extinction as mentioned above.
\citet{Fitzpatrick-2009} also demonstrated that
the extrapolations of the power-law fits from shorter wavelengths
(0.6--0.95~{$\mu$m}) are inconsistent with the near-IR ($\JHKs$) data points.
Including data sets in the $I$ band or shorter
wavelengths always lead to a smaller $\alpha$ value.
On the other hand,
\citet{Fitzpatrick-2009} also investigated the form (their Equation 5) 
\[
\frac{E_{\lambda -V}}{E_{B-V}} = [0.349+2.087 R_V] \frac{1}{1+(\lambda /0.507)^\alpha} -R_V
\]
to represent the extinction law with two free parameters,
$\alpha$ and $R_V$. This is not a power-law,
but it is successful in reproducing the extinction law
in the optical wavelengths longer than the $V$ band,
as well as in the near IR.
They demonstrated that, unlike the UV case \citep{Fitzpatrick-2007},
the optical--near IR extinction law may be well
represented by a formula with only a few parameters
(based on the data towards a limited number of sightlines),
but establishing such a formula needs further investigation.

\subsection{The interstellar extinction law: mid-IR}
\label{sec:ExtMIR}

The mid-IR extinction law had been scarcely studied
before the launch of {\it Spitzer Space Telescope} in 2003.
Some earlier works \citep{Landini-1984,Rieke-1985,Martin-1990} suggested
that a power-law extends from the near-IR to the $M$ and $N$ bands, whereas
others \citep{Lutz-1996,Lutz-1999} found shallower extinction law
at $\lambda > 3$~{$\mu$m}.

Precise and wide-field studies of the mid-IR extinction law
had become very active with {\it Spitzer}.
The mid-IR camera, InfraRed Array Camera (IRAC), can collect
images in four bands ([3.6], [4.5], [5.8], and [8,0]), and
deep mid-IR photometry for strongly obscured objects has become possible
thanks to its high sensitivity.
\citet{Indebetouw-2005} pioneered in the extinction studies with the {\it Spitzer}.
They observed the giant H~II region RCW 49
and two lines of sight, $l = 42^{\circ}$ and $284^{\circ}$,
and derived a flat extinction law at 3--8~{$\mu$m}.
A similar flat behavior has been supported by observations of various
targets and regions:
dark clouds and star forming regions
\citep{Flaherty-2007,RomanZunig-2007,Cambresy-2011,Ascenso-2013,Wang-2013},
the GC and Galactic bulge 
\citep{Nishiyama-2009,Chen-2013},
and different locations of Galactic plane 
\citep{Gao-2009,Xue-2016}.

The flattening of the extinction law for {$\sim$}3--8~{$\mu$m}
has also been confirmed by observations with other instruments.
For example, 
\citet{Jiang-2006} found $A_{7~{\mu{\rm m}}}/\AKs$ values larger
than suggested by the power-law extension.
\citet{Schodel-2010} found a clear change of the power-law index $\alpha$
from the $H$ to $L^\prime$ bands toward the GC:
$\alpha = 2.21 \pm 0.24$ inferred from
data sets in $H$ and $\Ks$ bands, and $1.34 \pm 0.29$ from those in $K$ 
and $L^\prime$ bands.
\citet{Fritz-2011} confirmed the flattening at $\lambda \gtrsim 3$~{$\mu$m}
toward the GC,
using 18 hydrogen emission lines at $1 < \lambda < 8$~{$\mu$m}.
The extinction law obtained by \citet{Davenport-2014}
shows a spatial variation but also shows a flattening in
the {\it WISE} $W1$~(3.35~{$\mu$}m) and $W2$~(4.46~{$\mu$m}) bands.

\begin{figure}
  \includegraphics[width=\textwidth]{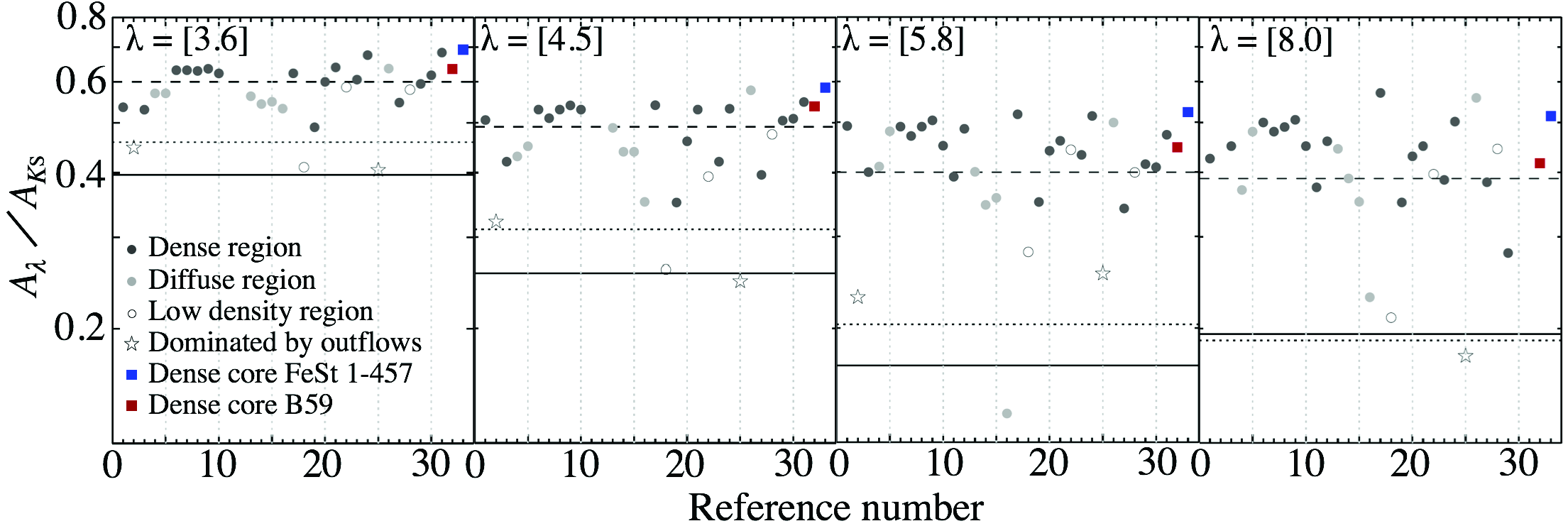}
\caption{
Comparison of $A_{\lambda}/\AKs$ for 
$\lambda = [3.6]$, $[4.5]$, $[5.0]$, and $[8.0]$
(from left to right),
adapted from \citet[][Fig.~6]{Ascenso-2013}.
The $A_{\lambda}/\AKs$ value by dust models are indicated by horizontal lines:
solid line and dotted line for the model by \citet{Draine-2003a,Draine-2003b} 
for $R_V = 3.1$  and $R_V = 5.5$, respectively, and  
dashed line for the model by \citet{Weingartner-2001} for $R_V = 5.5$. 
See \citet{Ascenso-2013} for references and more detail.
}
\label{fig:13AscensoFig6}
\end{figure}

\citet{Ascenso-2013}
compiled the $A_{\lambda}/\AKs$ values for the mid-IR range from the literature
between 1999 and 2013 (Fig.~\ref{fig:13AscensoFig6}).
Most of the studies use {\it Spitzer}/IRAC data sets, 
but those based on other data sets at similar wavelengths are also included
\citep{Bertoldi-1999,Lutz-1999,Fritz-2011,Olofsson-2011}.
If the mid-IR extinction law is an extension of the near-IR power-law,
Fig.~\ref{fig:13AscensoFig6} should show a decrease of $A_{\lambda}/\AKs$
from the left to right panels
(see the solid lines, $R_V = 3.1$, \citealt{Draine-2003a,Draine-2003b}).
However, the change of $A_{\lambda}/\AKs$
from [3.6] to [8.0] is rather small.
This together with the above results allows us to conclude that
the extinction law at $3 < \lambda < 8$~{$\mu$m} is not a simple extrapolation
of a power-law from shorter wavelengths
and the extinction is most often larger than expected from the power-law.

A color-excess ratio at mid-IR wavelengths, 
$E_{[3.6] - [4.5]} / E_{[4.5] - [5.8]}$,
was investigated by \citet{Flaherty-2007} for five star forming regions.
They found that the ratio toward Ophiuchus, $1.25 \pm 013$,
is significantly smaller than those toward the other regions,
which range from 2.49 to 2.61.
This result strongly suggests variation of the mid-IR extinction law
among dense star-forming regions.

\begin{figure}
  \includegraphics[width=0.65\textwidth]{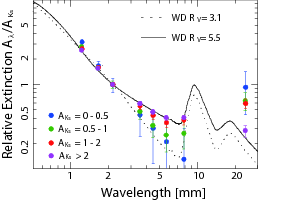}
\caption{Near-IR and mid-IR extinction law toward the Ophiuchus molecular cloud 
\citep{Chapman-2009b}.
Four observational results for different $\AKs$ ranges are shown.
The model extinction laws of \citet{Weingartner-2001} for $R_V = 3.1$ and $5.5$
are also plotted as solid and dotted lines, respectively.
Adapted from \citet[][Fig.~16]{Chapman-2009b}.}
\label{fig:Chapman09Fig16}
\end{figure}

An interesting result for the variation of the mid-IR extinction law 
was obtained by \citet[][reference \#18--21 in Fig.~\ref{fig:13AscensoFig6}]{Chapman-2009b}.
They explored changes of the near-IR and mid-IR extinction laws
in three molecular clouds, Ophiuchus, Perseus, and Serpens.
They found that 
the extinction law tends to be flatter along lines of sight
with stronger extinction (Fig.~\ref{fig:Chapman09Fig16}).
For $\AKs < 0.5$, the data points lie almost on 
top of the power-law extension from shorter wavelengths.
On the other hand, the extinction laws become as flat as
those found in other studies for $\AKs \gtrsim 1$.

Using red giants and RC giants,
\citet{Gao-2009} determined the color-excess ratios,
$E_{\Ks-\lambda}/E_{J-\Ks}$ and $E_{\Ks-\lambda}/\EHKs$,
for the four IRAC bands ($\lambda = $ [3.6], [4.5], [5.8], [8.0]).
By assuming $A_J/\AKs = 2.52$ and $A_H/\AKs = 1.56$,
they derived average $A_{\lambda}/\AKs$ values
for 131 fields along the Galactic plane.
The mean extinction ratio is consistent with the previous studies
but shows a slightly flatter wavelength dependence.
They also found a variation of $A_{\lambda}/\AKs$
along the Galactic plane.
In the case of red giants, $A_{\lambda}/\AKs$ varies
in the range from 0.07 to 0.16 for the four bands,
with a typical uncertainty of 0.001.
They also suggest systematic variations of $A_{\lambda}/\AKs$ 
with Galactic longitude.
$A_{\lambda}/\AKs$ seems to be smaller at around the Galactic longitudes
corresponding to the tangent positions of spiral arms,
which can be explained by grain growth in the arms.

\begin{figure}
  \begin{center}
    \includegraphics[width=\textwidth]{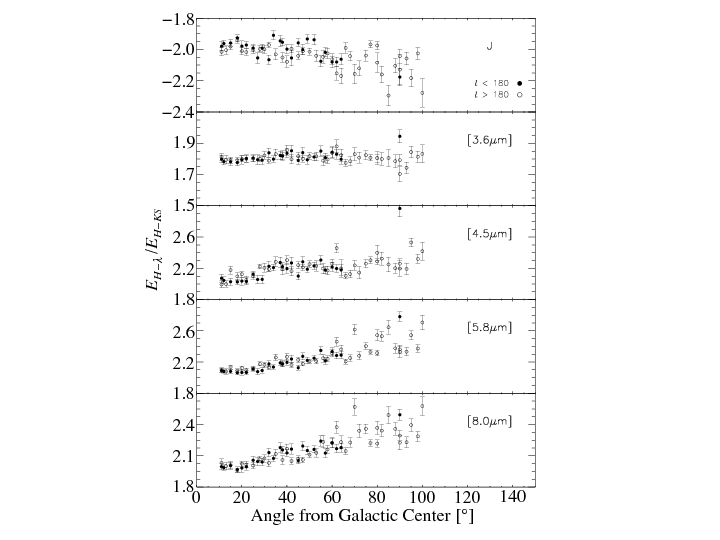}
    \caption{
      $E_{H-\lambda}/E_{H-K_{\rm s}}$ as a function of angle from the GC
      \citep{Zasowski-2009}.
      The filled and open circles are data points 
      for $l <180^{\circ}$ and $l > 180^{\circ}$, respectively.
      Adapted from \citet[][Fig.~5]{Zasowski-2009}.}
    \label{fig:09ZasowskiFig5}
  \end{center}
\end{figure}

\citet{Zasowski-2009} showed that color-excess ratios
vary as a function of Galactic longitude.
They used RC giants as tracers of extinction,
along the Galactic plane ($| b | \lesssim 1.0^{\circ} - 1.5^{\circ}$) 
in the first ($10^{\circ} < l < 65^{\circ}$),
and the third to fourth quadrants ($-105^{\circ} < l < -10^{\circ}$).
Fig.~\ref{fig:09ZasowskiFig5} plots the observed color-excess ratios,
$E_{H-\lambda}/E_{H-K_{\rm s}}$.
Here the color-excess ratios are not converted to 
the ratio of the amount of extinction, such as $A_{\lambda}/\AKs$,
allowing us to compare the extinction law 
without uncertainties from either the extinction ratio
or the ratio of total-to-selective extinction.
It is shown that the wavelength dependence of the extinction gets steeper
with increasing Galactic longitude,
while the ratio $E_{H-[3.6]}/\EHKs$ is nearly constant.
The trend of the $E_{H-\lambda}/E_{H-K_{\rm s}}$ ratios looks
identical between positive and negative Galactic longitudes.

The mid-IR extinction law has been found to be
flatter in denser molecular cloud regions
\citep[e.g.][]{Chapman-2009b,McClure-2009}.
The flattening can be interpreted as a result of grain growth
from either the coagulation of grains or
the accumulation of refractory ice mantles.
The steeper extinction law found by \citet{Zasowski-2009}
for larger angles from the GC
indicates a steady decrease of the mean dust grain size in the outer Galaxy.

\citet{Chen-2013} investigated the variation of the extinction ratio
$A_{\lambda} / \AKs$ for the four {\it Spitzer}/IRAC bands.
They combined the IRAC data with GLIMPSE-II and VVV data sets
for the Galactic bulge,
$-10^{\circ} < l < +10^{\circ}$ and $-2^{\circ} < b < +2^{\circ}$.
Their main purpose was to construct a 3D extinction map
by comparing observed photometric data (2MASS, VVV, GLIMPSE) 
with the Besan\c{c}con Galactic model.
Since they made extinction maps using a color excess $E_{\lambda -\Ks}$,
they can derive the ratio of the extinction, $A_{\lambda} / \AKs$,
using the color-excess ratios, $E_{\Ks - \lambda}/E_{J-\Ks}$
when the near-IR ratio $A_J/\AKs$ is assumed.
They found variations in $A_{\lambda} / \AKs$, 
and the variations show similar behaviors in all the IRAC bands.
As a function of Galactic latitude, there is a strong peak
toward the Galactic plane.
This suggests a stronger variation of $A_{\lambda} / \AKs$
along the Galactic latitude than along the Galactic longitude
within their target region toward the bulge.
The strongest variation is found at 5.8~{$\mu$m},
while the variation at 8.0~{$\mu$}m is less pronounced.
The extinction at around 5.8~{$\mu$}m appears as
continuum extinction.
The extinction at around 8.0~{$\mu$}m, in contrast, is largely attributed to
the 9.7~{$\mu$}m silicate absorption feature.
The smaller variation of the extinction at the latter wavelength suggests that
the ratio of the silicate absorption to the total extinction does not
vary so much along the Galactic latitude.

Note that results on the variations in the ratio of the amount of extinctions, 
such as $A_{\lambda}/\AKs$ at mid-IR wavelengths,
may depend not only on the mid-IR extinction law itself
but also on the variation in the near-IR extinction law.
In many cases, what we can determine directly from observations
is a color-excess ratio, 
e.g.\  $E_{K_{\rm s} - \lambda}/E_{H-K_{\rm s}}$.
Such a ratio can be translated into $A_{\lambda}/\AKs$  as 
\begin{eqnarray}
\frac{A_{\lambda}}{\AKs} 
&=& \frac{E_{\Ks - \lambda}}{E_{H-\Ks}} \biggl( \frac{A_H}{A_\Ks} - 1 \biggr) + 1 \nonumber \\
&=& \frac{E_{\Ks - \lambda}}{E_{H-\Ks}} \frac{E_{H-\Ks}}{\AKs} + 1. \nonumber
\end{eqnarray}
We thus need either a ratio of the amount of extinction
(in this case, $A_H/\AKs$)
or the ratio of the total-to-selective extinction ($\AKEHK$). 
These values are, however, usually difficult to determine directly
from observations. As shown in Fig.~10 in \citet{Nishiyama-2009},
the variation of the ratio of the amount of extinction at
mid-IR wavelengths,  $A_{\lambda}/\AKs$, 
may be attributed to differences in the near-IR law, $A_H/\AKs$.
Using color-excess ratios allows one to investigate 
the extinction law without obtaining or assuming 
the ratios with amount(s) of extinction involved
(e.g.\  $A_H/\AKs$ and $\AKEHK$) as demonstrated, e.g.\  by
\citet{Zasowski-2009}. The color-excess ratios depend much less
on the filter wavelength than the power-law index.
Some observational approaches in which parameters like $\AKEHK$ 
are directly measured will be discussed in \S\ref{sec:Remarks}.

It should be borne in mind that 
the mid-IR wavelength region contains various emission and absorption features.
The most prominent one is the 9.7~{$\mu$m} silicate absorption feature.
When we observe the line of sight toward the GC,
absorption features such as H$_2$O, CO$_2$, NH$_3$, and CH$_4$ are seen
from {$\sim$}3 to {$\sim$}8~{$\mu$}m \citep{Chiar-2000},
which are in the range of the {\it Spitzer}/IRAC bands.
The mid-IR hosts the Polycyclic Aromatic Hydrocarbon (PAH)
emission features in the range of 3--20~{$\mu$}m.
For wavelengths longer than {$\sim$}8~{$\mu$}m,
the extinction may also be underestimated because of the possible presence
of dust emission.
The late-type giants may have thin shell(s) of silicate dust,
and they may exhibit the 9.7~{$\mu$}m silicate emission.
In addition, the observation could be affected by dust thermal emission.
Although the extinction is near a minimum in the wavelength range of 
3--8~{$\mu$}m,
it is necessary to take into consideration the potential impact
of the emission and absorption features described above
when we try to derive the mid-IR extinction law.

In summary, the extinction law in the mid-IR is still 
less well established compared to that at optical and the near-IR
mainly because the opportunities of obtaining precise photometric measurements 
are limited. However, the extinction at 3--4~{$\mu$m}
is about half of that in the $K$ band, and therefore 
the effects of the remaining uncertainties can be also small.
This encourages one to further investigate the extinction law in the mid-IR 
and applications to various tracers.

\section{Cepheids in the Galactic Disk}
\label{sec:CepDisk}

Cepheids have already played a prominent role 
in our understanding of the Galactic disk's structure and evolution: 
their three-dimensional distribution in the disk 
delineates the spiral pattern \citep{Joy-1939,Majaess-2009,Dambis-2015}; 
their kinematics provides firm estimates of the Oort constants 
\citep{Kovacs-1990,Metzger-1991,Feast-1997,Pont-1997,Metzger-1998},
while their chemical abundances can be used for
accurate measurements of the radial abundance gradient across the disk  
\citep{Genovali-2014,Genovali-2015,Andrievsky-2016,daSilva-2016}. 

The sample of known Cepheids in the Galaxy is not only limited
({$\sim$}600) but also strongly biased.
Over 10,000 Cepheids are expected to populate the Galaxy \citep{Windmark-2011},
but only {$\sim$}5~\% (predominately nearby ones) have been identified so far.  
The currently known Cepheids are concentrated mostly within a few kpc
around the Sun since they have been discovered based on optical surveys
which are hampered by the high interstellar extinction toward
the Galactic plane.
Moreover, such surveys were performed by using different
instruments and photometric techniques, producing strongly inhomogeneous
data sets which can be hardly merged for a statistically coherent analysis.  

Fortunately, the Cepheid sample is improving rapidly,
especially thanks to the ongoing effort from large-sky time-domain surveys.
Near-IR variability surveys have produced small catalogs to date,
but they have discovered Cepheids in the most obscured  
regions of the Galaxy, challenging the current observational bias. 
In particular, near-IR surveys using the Infrared Survey Facility (IRSF)
in South Africa toward the Galactic plane 
led to the discovery of four new Cepheids  
in the Nuclear Stellar Disk \citep{Matsunaga-2011,Matsunaga-2015}, 
three at 3--5~kpc from the GC \citep{Tanioka-2017},
13 beyond the GC \citep{Matsunaga-2016}, and five
parallel to the Galactic central bar (Inno et al.\  in prep).
Near-IR data collected by the ESO public survey 
VVV \citep{Minniti-2010} have also been used to discover 
37 Cepheids located beyond the GC
\citep{Dekany-2015a,Dekany-2015b}.
With optical surveys, on the other hand, the Optical Gravitational
Lensing Experiment (OGLE)
has recently announced the discovery of $\sim$2,000
Cepheids found in the Southern disk \citep{Udalski-2017}
in addition to new Cepheids in less obscured regions
toward the Galactic bulge \citep{Soszynski-2017}. 
The Pan-STARRS1 3$\pi$ survey in $grizy$ bands also has a great potential
for identifying 
a large number of new Cepheids in the northern hemisphere
and the exploitation of its data set is pioneering for future
all-sky photometric surveys (e.g.\ LSST).
In the near future, {\it Gaia} will
allow us to discover a large number of Cepheids and other variables
across the sky. Its limiting magnitude is deep enough to detect
Cepheids to some distances, roughly a few kpc toward the inner Galaxy
and more in the outer Galaxy, even in reddened regions of the Galactic disk,
although it cannot reach the GC and beyond \citep{Windmark-2011}. 

While these surveys are widening the census of Galactic Cepheids,
they (even the {\it Gaia}'s all-sky survey) are still characterized
by unknown selection functions, given the uncertainty in the extinction  
throughout the Galactic disk and the different cuts in RA, Dec, magnitudes etc. 
For example, \citet{Matsunaga-2016} discussed a gap between
the optical and IR surveys toward the bulge. Moreover,
the interstellar extinction not only limits the completeness of Cepheid surveys
but also introduces large uncertainties on the Cepheids' distances.
A relatively small error in the extinction law can lead
to significant errors in distances to Cepheids in obscured regions
even when IR data are used for distance estimation.
A good example of such a problem is found by \citet{Matsunaga-2016}
who discussed conflicting estimates for distances to new Cepheids,
which are more distant than the GC, in their survey
and to Cepheids reported by \citet{Dekany-2015b}.
Readers are also referred to \citet{Matsunaga-2017} about
these Cepheids as well as general discussions on the effects of
interstellar extinction in the context of variable stars in the disk.

Although \citet{Matsunaga-2016} found that the extinction law toward
the bulge should be close to the one obtained by
\citet[][cf.\  \citealt{AlonsoGarcia-2017,Hosek-2018}]{Nishiyama-2006},
the situation may be different
for the Galactic disk due to the spatial variation as we discussed
in \S\ref{sec:Ext}. Such a spatial variation 
creates a very challenging problem for drawing 
a map across a wide range of the disk.
For example, \citet{Tanioka-2017} considered a large range of $\AKEHK$ values,
from 1.44 \citep{Nishiyama-2006} to 1.83 \citep{Cardelli-1989},
toward the inner disk, $l=+20^\circ$ and $+30^\circ$.
This large uncertainty in the extinction law prevented them from
obtaining an accurate distance to moderately obscured Cepheids;
the error in the distance modulus is as large as 0.8~mag
with $\EHKs \sim 1.8$~mag. Such large errors limit
the use of Cepheids, and of any other tracers whose distances are
obtained based on photometric data, to draw
a map of stellar distribution in the disk.
The characterization of the extinction law is an urgent task
to exploit the outcomes of the ongoing large-scale surveys.
Combining more multi-band magnitudes including mid-IR ones
(partly done in \citealt{Tanioka-2017} using 3.6~{$\mu$m} data)
would reduce such errors once the extinction coefficients
in the mid-IR bands are investigated as well as in the near-IR $\JHKs$ bands.


\section{Open Clusters}
\label{sec:OC}

A collection of 2,167 open clusters with relatively reliable parameters
is found in \citet[][see the version updated in January 2016]{Dias-2002}, which
encompasses an update to the previously published catalogs of
\citet{Lynga-1987} and \citet{Mermilliod-1995}. \citet{Kharchenko-2013}
derived open cluster parameters for 
3,007 clusters based on 2MASS $\JHKs$ photometry and PPMXL proper
motions. The samples contained in both of these catalogs are complete
out to 1.8~kpc. With new near-IR data, such as the VVV's,
hundreds of open clusters that are more distant and heavily obscured 
have been found \citep{Borissova-2011,Barba-2015}.  

Since open clusters can provide valuable information on age,
distance, and extinction based on the isochrone fitting,
they can serve as vital probes of the Galactic disk.
Interesting results of such applications
are found in various papers from long ago to very recent
\citep[][and references therein]{Reddy-2016}.
In the past few years, a number of clusters containing red
supergiants have been found in the interface of the Galactic (long)
bar and the inner disk
\citep[e.g.][]{Davies-2009,Davies-2012}.
The formation of these clusters are suggested to be 
triggered by the interaction between the bar and the disk. 
Characterizing the parameters of the clusters is, however, quite expensive.
Although we already have a large number of samples which are ready
to draw a comprehensive picture of the Galactic structure
in the solar neighborhood \citep{Joshi-2005,Joshi-2016},
samples beyond several kpc are still being discovered and a continuous effort 
needs to be made as to follow-up observations. An outstanding effort in
the spectroscopic approach is, for example, done by the Gaia-ESO survey
\citep{Magrini-2017}, whose samples are still rather limited compared to
the potential sample size of open clusters.
In this review, we focus on Cepheids and contact binaries in open clusters,
these two kinds of stellar tracers can give distances by themselves 
and those in the clusters are useful for establishing 
the distance scales (e.g.\  mutual calibration of different
distance indicators).

\subsection{Cepheids in open clusters}
\label{sec:CepOC}

The method of estimating distances to open clusters based on
the main-sequence fitting also provides us with an opportunity
to calibrate (or cross-validate) the PL relations of Cepheids.
Since the first Cepheids were found in the open clusters NGC~6087 and
M~25 by \citet{Irwin-1955}, many researchers have contributed to this
field \citep[e.g.][]{vandenBergh-1957,Tsarevsky-1966,Turner-1986,Turner-1993,Turner-2010,Baumgardt-2000,Hoyle-2003,Tammann-2003,An-2007,Majaess-2008}.
In particular, \citet{Tammann-2003} derived $BVI$-band Cepheid PL
relations based on 25 cluster Cepheids, whereas
\citet{Anderson-2013} identified five new
cluster Cepheids based on an eight-dimensional parameter search and
obtained a $V$-band PL relation for 18 cluster Cepheids;
the membership confidence is evaluated for each open cluster--Cepheid pair 
based on spatial position (two-dimensional angular distance and parallax),
kinematics (radial velocity and two-dimensional proper motion),
and population (age and metallicity).
Recently, \citet{Chen-2015,Chen-2017} found nine new cluster Cepheids
based on the near-IR 2MASS and VVV catalogs. 
To date, approximately 40 cluster Cepheids have been found and confirmed,
of which 30 can be used to derive PL relations in various photometric filters.
In the last decade, more and more attention has been paid to the near-IR or
even to mid-IR PL relations, since at these longer wavelengths, the PL
relations have much smaller intrinsic scatter and are less affected by
interstellar extinction. Based on 31 cluster Cepheids,
\citet{Chen-2015,Chen-2017} derived new $\JHKs$ PL relations, which represented
improvements of up to 40~\% compared with previous determinations.
These relations were used by \citet{Wang-2017} to derive mid-IR PL relations
of hundreds of Galactic Cepheids. 

\begin{figure}[h!]
\centering
\includegraphics[width=95mm]{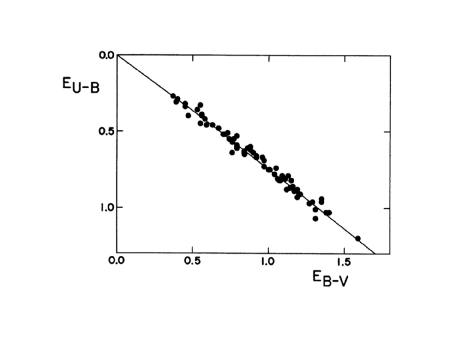}
\caption{69 OB-type stars in NGC~6611 and Sgr OB1 were used to derive
  the color-excess relation. The black line is the best-fitting line,
  $E_{U-B}/E_{B-V}=0.721+0.025 \times E_{B-V}$. This is Fig.~1 of
  \citet{Turner-1989}. }\label{fig:xdf1}
\end{figure}

\begin{figure}[h!]
\centering
\includegraphics[width=95mm]{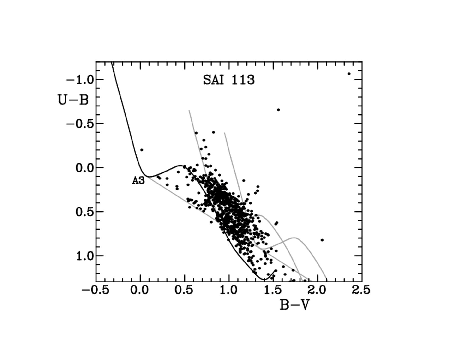}
\caption{$UBV$ color--color diagram for stars within 6~arcmin of the
  center of SAI~113. The black curve is the intrinsic curve, while the
  two gray curves are for color excesses of $E_{B-V}=0.88$ and $1.28$~mag
  The gray line shows the A3 `kink' stars for different
  extinction values; this provides information on extinction law,
  $E_{U-B}/E_{B-V}=0.64$. This figure is the top panel of Fig.~6 from
  \citet{Carraro-2017}. }\label{fig:xdf2}
\end{figure}

As already discussed in \S\ref{sec:CepDisk},
the uncertainty in the extinction law
has a large impact on distances to objects obscured
by severe interstellar extinction.
Here, open clusters can play an important role in disentangling this problem;
they can provide an independent reddening value as well as a distance.
Once the spectral types of cluster members are obtained,
their color excesses can be estimated based on
the difference between apparent and intrinsic colors. Based on two
or more color-excess values, a more accurate extinction ratio can be
estimated for stars within each particular cluster.
If spectroscopic observations of
most clusters are difficult to obtain, the optimal reddening value can
instead be evaluated based on application of the isochrone fitting
method. \citet{Turner-1989} used $UBV$ photometry of six open clusters or
associations to study the reddening relation, $E_{U-B}/E_{B-V}=X+Y
\cdot E_{B-V}$. He found that the second-order term is almost
negligible, and that $X$ ranges from 0.62 to 0.80. 
This method allows one to estimate the reddening ratio for each open cluster
independently based on detailed analysis of its cluster members.
Fig.~\ref{fig:xdf1} shows the reddening relation 
derived for early-type stars in the clusters NGC~6611 and Sgr~OB1. 
Fig.~\ref{fig:xdf2} presents another example which can be used to
derive the reddening ratio based on the color--color diagram.
The $\JHKs$ color--color diagram also shows a `kink' feature as the $UBV$
color--color diagram \citep{Turner-2011} , so this is a potential approach
for investigating variations in the near-IR extinction law.
\citet{Turner-2011} provided an empirical 
$\JHKs$ magnitudes and colors of ZAMS in his Table~1, and applied them to five
open clusters. \citet{Chen-2015} also used the $\JHKs$ color--color diagram
to study the near-IR extinction for 17 open clusters. 
Most of previously studied clusters have relatively small color excesses,
e.g.~$0.06\leq E_{J-H} \leq 0.36$ in \citet{Chen-2015}.
This limits the ability to constrain the extinction law.
Hopefully, with new near-IR instruments and projects such
as the VVV survey, more heavily reddened open clusters can be found
and investigated. More individual near-IR extinction laws
for individual open clusters may be obtained in the next few years.

A particularly interesting cluster in our context is VdBH~222
in which a long-period Cepheid ($P\sim23.325$ days)
was reported by \citet{Clark-2015}. Prior to their work,
two possible kinematic distances were suggested by \citet{Marco-2014}, 
{$\sim$}6~kpc and {$\sim$}10~kpc, based on the radial velocity of the cluster,
but they suggested that the distance of 10~kpc is more consistent with
Galactic structure. As regards the age of VdBH~222 (based on
the isochrone fitting), the distance of 10~kpc would suggest 14~Myr,
while the distance of 6~kpc corresponds to 20~Myr.
The Cepheid provides two constraints, an independent and direct distance 
to the cluster and a constraint on its age.
Based on the Cepheid PL relation and the
best-fitting reddening value from \citet{Marco-2014}, a distance of
$5.8_{-0.7}^{+0.8}$~kpc was estimated, while an age of
$24.5_{-4.1}^{+5.0}$~Myr was determined by adoption of the Cepheid
period--age relation of \citet{Bono-2005}. This Cepheid is thus
consistent with a comparably high isochrone age and a small distance.
The distance of 6~kpc places VdBH~222 in the 3-kpc inner arm,  
which was unexpected since star formation was considered to be
suppressed in this arm \citep[see the discussions in][]{Clark-2015}.
The properties of VdBH~222 also suggest
that ${\sim}10^4 M_\odot$ of stars have been formed in the last
{$\sim$}20~Myr, which provides further evidence of active star formation
at the interface of the bar and the disk.
As regards the problem on the extinction law, the
color excess of the cluster, $E_{J-\Ks} \sim 1.1$, is larger than those of
other clusters investigated so far. This would allow us to give a more
tight constraint on the extinction law, although such a study has not been done.

\subsection{Contact binaries in open clusters}
\label{sec:CBOC}

For contact binaries, one can expect PL (or PLC) relations as follows.
Let us start with the basic equations which govern contact binaries:
\begin{eqnarray}
m_1 +m_2 \propto a^3 / P^2 \nonumber \\
L \propto \Teff ^4 S ,  \nonumber
\end{eqnarray}
where $m_1$ and $m_2$ are masses of the two components,
$a$ is the semi-major axis, $P$ is the orbital period,
$L$ is the total luminosity of the system, and $S$ is
the total projected area seen by an observer.
In each contact binary system,
material transfer is ongoing in their common envelopes, and
both components have filled their Roche lobes.
The two components typically have the same temperature, $\Teff$. 
For a very simple case
of two stars with the identical mass, $m$, and radius, 
then $a$ gets close to the radius of each component
and $S$ is proportional to $a^2$ at the maximum light.
In such a case, the above two relations can be combined to give
\begin{eqnarray}
L_{\rm max} \propto \Teff^4 m^{2/3} (1+q)^{2/3} P^{4/3} , \nonumber 
\end{eqnarray}
where $L_{\rm max}$ is the maximum luminosity and $q$ is the mass ratio,
$m_2/m_1$, but $q=1$ for the simplest case.
Main-sequence stars follow a mass-temperature relation,
$\log m \sim \gamma \log \Teff$ ($\gamma \sim 4/3$, \citealt{Eker-2015}),  
which leads to a period--luminosity--temperature relation
(or to a PLC relation),
\begin{eqnarray}
\log L_{\rm max} = (4/3)\log P + [4 +(2/3)\gamma] \log \Teff + {\rm const}. \nonumber 
\end{eqnarray}
Moreover, $\Teff$ is also correlated with 
mass and radius in case of main-sequence stars \citep{Eker-2015},
and the period--luminosity--temperature relation can be reduced to 
a PL relation for the simple case being discussed.
In realistic case of contact binaries, it has long been known
since a seminal work by \citet{Eggen-1967} that
they show a correlation between period and color (and thus temperature)
although \citet{Eggen-1967} also suggested that the correlation is
blurred by the degree of contact.

The above explanation is of course oversimplified.
First, two components are no longer spheres but
fill the Roche equipotentials.
Furthermore, the mass ratio $q$ and
the degree of contact affect the relation between
the total area $S$ and the semi-major axis $a$, and
the gravity-darkening effect \citep{Lucy-1967} and
other effects need to be taken into account.
Nevertheless, at least a large fraction of contact binary systems
show tight correlations between the physical parameters,
and are governed by simple relations as demonstrated here.
In his classical textbook, \citet{Kopal-1959} calculated
the geometry of contact binary systems, in particular those with
both components filling the Roche lobes.
Table~3-3, in Section~III.4 in the book, lists the radii of both components,
$r_1$ and $r_2$, in units of the major axis ($2a$)
for various mass ratios between 0 and 1.
The simple projected area, $S$, at the maximum can be estimated as
$\pi ( r_1^2 + r_2^2 )$, and it is nearly constant for $0.2 \leq q \leq 1$
within $\pm 10$~\%
while it significantly increases towards $q=0$ at lower $q$.
Systems with small $q$ tend to be missed, or can be ignored if necessary,
because of their small amplitudes; see the relation between
$q$ and amplitude in \citet{Rucinski-2001}.
Furthermore, as discussed by \citet{Eggleton-1983},
$P(\rho)^{1/2}$ shows only a little
variation over a wide range of $q$.
This constraint is analogous to
the period--mean density relation of pulsating stars.
Combined with the limited distribution of 
main-sequence stars and contact binaries themselves \citep{Eyer-2008}
on the HR diagram, 
like the instability strip for Cepheids, one can expect
a PL (or at least a PLC) relation.
For more detailed calculations, readers are referred to, e.g., 
\citet{vanHamme-1985}, \citet{Rucinski-1994,Rucinski-2004},
and \citet{Chen-2016b}.

Contact binaries are divided into early and late types based
on their spectral types; 
the two groups have different temperatures of the two stars in each system,
and their orbital periods also tend to be different (longer for the former and shorter for the latter).
Late-type contact binaries are also called W Ursa Majoris (W UMa) type,
and most well-studied contact binaries belong to this type. 
Early-type contact binaries are relatively rare and poorly studied
compared with W UMa type. Contact binaries are still ubiquitous
in the Universe; their number is larger than those
of any other variable stars, without solar-like oscillation or
other tiny variations into account. Their density is approximately 0.2~\% in
the solar neighborhood and the Galactic bulge but may be slightly lower
in the Galactic plane, 0.1~\% \citep{Rucinski-2006}.
In intermediate-age open clusters, contact binaries can be
as abundant as 0.4~\%, but they are rare in young (100~Myr)
and old (10~Gyr) open clusters. Because they are abundant and 
can be easily identified, contact binaries are useful as
potential distance indicators.

To establish their PL relation, parallaxes and isochrone-fitting
distances to open clusters hosting contact binaries
are two independent distance estimates that can be used.
\citet{Rucinski-1994} used 18 W UMa binaries in open clusters to obtain
a $V$-band PLC relation, whereas
\citet{Rucinski-1997} obtained a $V$-band PLC relation,
$M_V=-4.30 \log P +3.28(B-V)_0+0.04, \sigma=0.17$,
based on 19 binaries with {\it Hipparcos} parallax
accuracy better than 0.25~mag in distance modulus.
The latter relation has particularly been used to anchor distances of contact binaries, 
although its color dependency is high. About one decade later,
\citet{Rucinski-2006} tried to reduce the PLC relation to a PL relation
and obtained a $V$-band PL relation, 
$M_V = -1.5 (\pm 0.8)-12.0 (\pm 2.0) \log P, \sigma=0.29$,
based on 17 W UMa binaries. This PL relation has larger
uncertainties in both coefficients, which limits its application.
Since then, with some questions on the reliability and the accuracy of
W UMa binaries, their application as a distance indicator hasn't been
active compared to classical distance indicators like Cepheids.

\begin{figure}[h!]
\centering
\includegraphics[width=95mm]{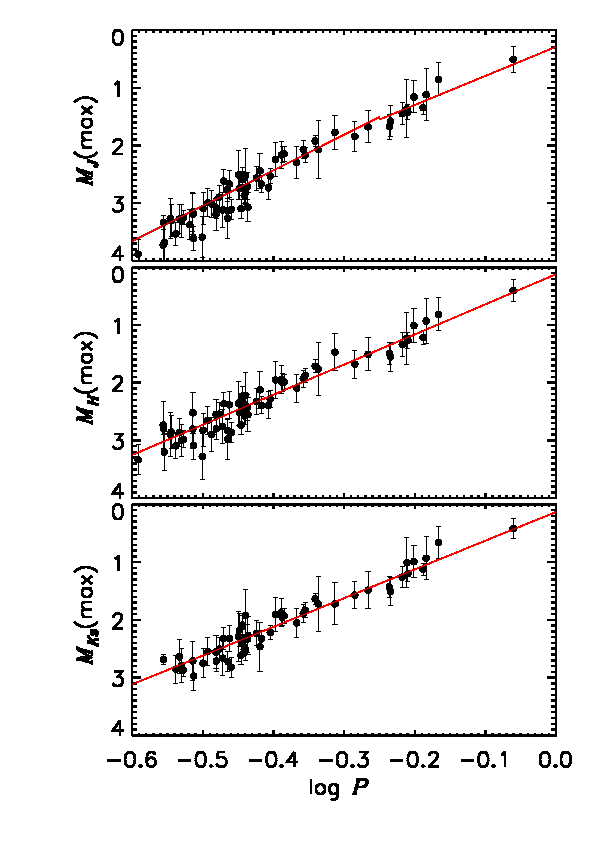}
\caption{46 contact binaries with open cluster or nearby moving-group distances and 20 contact binaries
  with high-accuracy {\it Hipparcos} parallaxes. The red line shows the
  best-fitting PL relations. Adopted from \citet[Fig.~1]{Chen-2016b}. 
  }\label{fig:xdf3}
\end{figure}

\begin{figure}[h!]
\centering
\includegraphics[width=95mm]{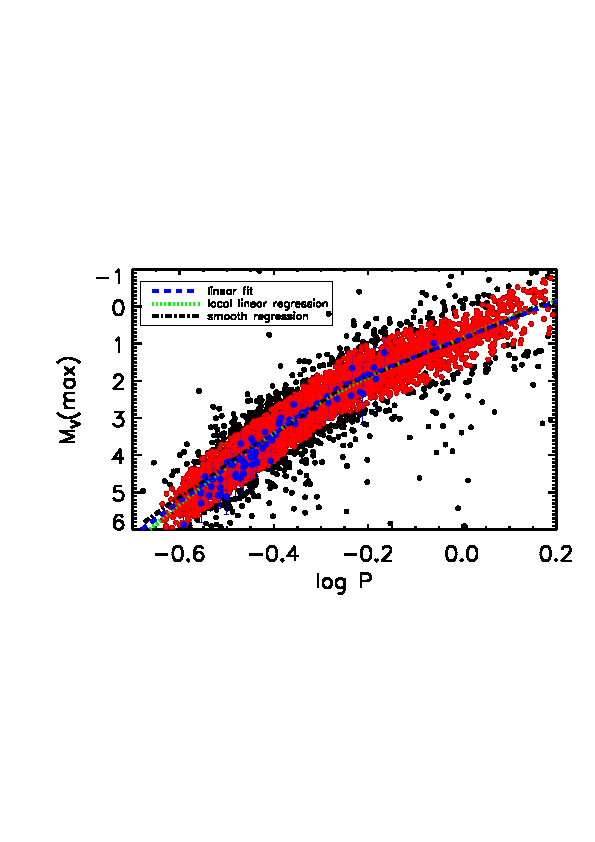}
\caption{$V$-band PL relation based on 6090 contact binaries (black dots). Red dots
  are stars in the 2$\sigma$ distribution. The different lines denote
  different fitting methods; the non-parametric fits agree with the
  two linear line fits. The early- and late-type contact binaries have different
  $V$-band PL relations with a boundary at $\log P=-0.25$ days. 
  Adopted from \citet[Fig.~5]{Chen-2016b}. }\label{fig:xdf4}
\end{figure}

\citet{Chen-2016a} discussed physical parameters of
contact binaries in NGC~188 in detail, and
suggested the presence of an accurate PL relation,
whose slope is shallower than that derived by \citet{Rucinski-2006},
considering five confirmed contact binaries
in NGC~188 combined with seven in two other clusters
(M~67 and Berkeley~39; see Fig.~7 in \citealt{Chen-2016a}).
Then, \citet{Chen-2016b} collected data for 66 contact binaries with
accurate independent distances and derived the near-IR PL relations,
\begin{equation} 
  \begin{aligned}
   M_{J_{\rm max}}^{\rm late}  &= -6.15~(\pm 0.13) \log P-0.03~(\pm 0.05)  &~({\rm for} \log P < -0.25), \nonumber \\
   M_{J_{\rm max}}^{\rm early} &= -5.04~(\pm 0.13) \log P+0.29~(\pm 0.05)  &~({\rm for} \log P > -0.25), \nonumber \\
   M_{H_{\rm max}}             &= -5.22~(\pm 0.12) \log P+0.12~(\pm 0.05), &~ \nonumber \\
   M_{K_{\rm s,max}}           &= -4.98~(\pm 0.12) \log P+0.13~(\pm 0.04), &~ \nonumber
   \end{aligned}
\end{equation}
with 1~$\sigma$ residual scatter of 0.08--0.09~mag (Fig.~\ref{fig:xdf3}).
These PL relations are fairly tight and can be used to measure distances
of individual objects with 5~\% accuracy.
The metallicity effect on these relations is also discussed
in \citet{Chen-2016b}, but it is unclear or not significantly larger than
the scatters of the relations \citep[see Fig.~6 in][]{Chen-2016b}.
In addition, to derive the $V$-band PL relation
(Fig.~\ref{fig:xdf4}),
they estimated the near-IR mean distance and mean extinction of
6090 contact binaries with $V$-band data.
\begin{equation} 
  \begin{aligned}
   & M_{V_{\rm max}} = -9.15~(\pm 0.12) \log P-0.23~(\pm 0.05) &~({\rm for} \log P < -0.25), \nonumber \\
   & M_{V_{\rm max}} = -4.95~(\pm 0.13) \log P+0.85~(\pm 0.02) &~({\rm for} \log P > -0.25). \nonumber
   \end{aligned}
\end{equation}
with residual scatter of 0.30 and 0.35~mag.
The early- and late-type contact binaries (separated at $\log P=-0.25$,
\citealt{Rucinski-2006})
follow different PL relations in the $V$ band,
while this discrepancy disappears in the $HK_{\rm s}$ bands.

Applying the $V$-band PL relation to 102 early-type contact binaries
in the Large Magellanic Cloud (LMC),
\citet{Chen-2016b} obtained an LMC distance modulus of
$(m-M)_0=18.41\pm0.20$~mag. This distance modulus is in accordance
with the current best distance to the LMC,
$(m-M)_0=18.49\pm0.09$~mag (summarized by \citealt{deGrijs-2014}).
More than 30,000 \citep{Chen-2016a}
contact binaries have been found thus far and
several hundred contact binaries are found every year in individual papers. 
With follow-up observations such as spectroscopy,
contact binaries can also provide age and kinematic information.
They are thus very powerful tracers of stellar populations in the Galaxy.

\section{Metallicity Gradient}
\label{sec:gradient}

When distances to many individual stellar tracers like Cepheids
and open clusters become available, 
an important scientific step is then to derive 
the chemical abundance patterns of these tracers in the disk.
In particular, the radial distributions of different elements
(e.g.~[Fe/H], [$\alpha$/H], etc.) 
are fundamental observables to constrain the recent disk chemical enrichment, 
since they can be directly compared with models 
\citep{Chiappini-1997,Matteucci-2001,Cescutti-2007,Kudritzki-2015}.
For example, the gradient is affected
by the radial migration (see the Introduction).
Constraining empirically the efficiency of the radial migration is difficult 
because stars in the disk quickly lose dynamical signatures of this process.
On the other hand,
the radial migration affects the chemical abundance patterns and
increases the observed metallicity dispersion at any given radius. 
Observational data of the abundance gradients for stars in a wide range of ages
would allow us to constrain the efficiency of orbit migration 
for blurring the initial abundance pattern.

\begin{figure}
  \includegraphics[width=\textwidth]{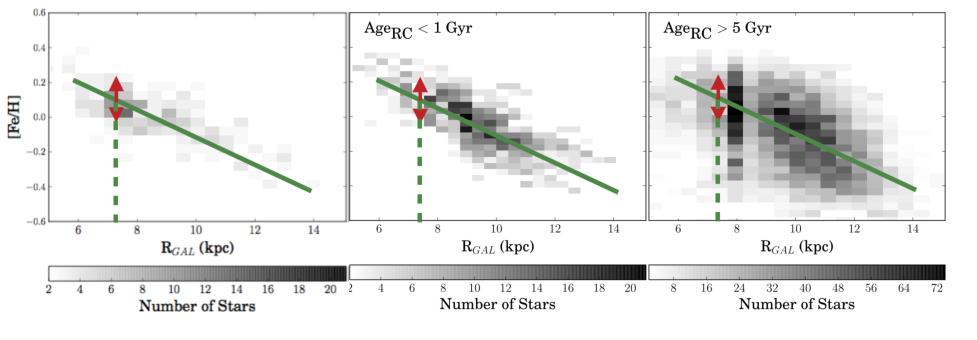}
\caption{Left panel: 2D histogram of the iron abundances of known
Galactic Cepheids \citep{Genovali-2015} as a function of
their Galactocentric radius ($\RGC$). 
The green solid line shows the best-fitting linear gradient,
which corresponds approximately to $-0.06$~dex~kpc$^{-1}$.  
The dashed line corresponds to a distance $\RGC=7$~kpc, where
the dispersion around the linear gradient:
$\pm \sigma_{{\rm [Fe/H]},R=7~{\rm kpc}}=0.1$~dex is indicated by
the red arrows, and is consistent with the nominal error in the abundance determination of individual stars. 
Middle panel: Similar but for RC stars younger than 1 Gyr, with iron abundances derived from APOGEE spectra \citep{Ness-2016}. 
The plotted lines are the same as in the left panel but they have been drawn here to guide the eye in comparing the dispersions. 
Left panel: Same as in the previous panels, but for older RC stars. 
Here the increase in $\sigma_{{\rm [Fe/H]},R=7~{\rm kpc}}$ is clearly detectable. Such increase in the dispersion is caused by the orbit migration in the Galactic disk.}
\label{fig:2}      
\end{figure}

Because of their young age, Cepheids should not have been mixed so much yet, 
thus their chemical abundances can be used to set a boundary condition
to the radial migration effects. 
In Fig.~\ref{fig:2}, we compare the iron abundances of Cepheids
in the sample of \citet{Genovali-2015} with those resulting from RC stars
derived on the basis of APOGEE spectra by \citet{Ness-2016}. 
On the basis of their content in $\alpha$-element abundances with respect to 
the iron abundances, \citet{Ness-2016} distinguished between
RC stars in the thick and thin disk. Moreover, they were also able to
determine the ages with an accuracy of better than 40~\% for the thin disk
sample. The left panel in Fig.~\ref{fig:2} shows the density distribution
of the iron abundance of Cepheids as a function of $\RGC$,
which follows a well established linear gradient
\citep{Genovali-2015},
while the middle and right panels show the distributions
of young and old RC stars, respectively \citep{Ness-2016}.
This plot clearly illustrates the effect of radial migration 
in the Galactic disk. In fact, it shows how the dispersion around the radial metallicity gradient
increases when we move from the younger (left) to the older stellar populations (right). 

The metallicity distribution of open clusters also shows
the radial gradient \citep{Friel-1993,Twarog-1997}.
Unlike Cepheids, there are
clusters with various ages from a few Myr to a few Gyr.
This makes it possible to study the time evolution of the metallicity gradient
\citep[e.g.][]{Tsujimoto-2010,Jacobson-2011}.
Compared to Cepheids, however, the number of open clusters with their
locations and metallicities well determined is still limited
(see recent progress, e.g., in \citealt{Magrini-2017}).
In addition, open clusters can be disrupted because of
internal (e.g.\  SN explosion) and external (e.g.\  Galactic potential)
effects, thus the observed galactic population may be biased.

An interesting new feature was found in recent IR surveys
revealing a rich group of massive stellar clusters
which hosts several red supergiants
\citep[e.g.][see also \S\ref{sec:OC}]{Davies-2012}.  
They seem to be preferentially found around the near- and far-side ends
of the Galactic bar although the global distribution of such massive clusters
in the inner part of the Galaxy remains to be concluded after more complete
surveys and detailed studies have been done. Surprisingly low metallicities
have been reported for some of these massive clusters 
\citep{Davies-2009,Origlia-2013,Origlia-2016},
around $-0.2$~dex or even lower in contrast to the high metallicities,
{$\sim$}0.3~dex at a similarly inner part of the disk
\citep[Fig.~3 and discussions therein]{Bono-2013}.
No firm scenario to explain such low metallicities has been proposed yet.
There is a large room of improvement for the near future both in
characterization of open clusters (especially their distances and
metallicities) and finding new clusters.
In addition, red supergiants in the inner Galaxy and even wider areas
will be useful as bright tracers of stellar populations
whether or not they belong to clusters
\citep[see e.g.][]{Messineo-2016,Messineo-2017}.

\begin{figure}
  \includegraphics[width=\textwidth]{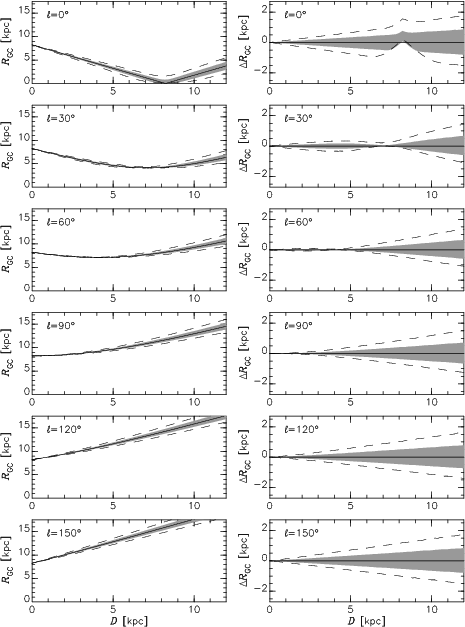}
\caption{
The Galactocentric distance ($\RGC$) and the deviation of estimated $\RGC$ 
are plotted against the distance, $D$, on the left and right side panels,
respectively, for seven lines of sight with different Galactic longitudes.
The solid curve in the middle indicates the true value, while
gray filled area and dashed curves indicate the ranges
of $1~\sigma$ errors corresponding
to the modulus errors in 0.15 and 0.3~mag.
\label{fig:calcRGC}}
\end{figure}

How is the study of the metallicity gradient affected
by uncertainties in the distance determination?
Fig.~\ref{fig:calcRGC} plots how the errors in distance are
translated to the errors in $\RGC$.
Note that we only consider the Galactic plane, i.e.~$b=0^\circ$, and 
it is assumed that the Sun is located at $\RGC = 8.3$~kpc \citep{Reid-2014a}.
The relation between the two errors depends on Galactic longitude.
For example, up to {$\sim$}5~kpc, a line of sight at 
$l \sim 60^\circ$ runs at almost the same $\RGC$ and thus
the error in $\RGC$ tends to be small. In the direction of $l=0^\circ$,
the kink corresponds to the location of the GC where
the error in $\RGC$ is always positive.
At around a distance of 10~kpc or beyond,
the sightline dependency of $\Delta \RGC$ becomes small,
and the errors of 0.15 and 0.30~mag in distance modulus correspond
to the $\RGC$ errors of approximately 0.4 and 1~kpc, respectively,
while these errors are less than half at the distance of 5~kpc.
The metallicity gradient is 0.06~dex~kpc$^{-1}$, for example,
in the case of Cepheids \citep{Genovali-2014} and the scatter of 
the metallicity of individual Cepheids
around the average trend is {$\sim$}0.1~dex. 
Detailed studies of substructures \citep[see e.g.][]{Genovali-2014}
would require a higher precision, {$\sim$}0.03~dex, at each $\RGC$.
Errors in $\RGC$ would introduce an artificial scatter
in the metallicity gradient. The Cepheids investigated by \citet{Genovali-2014} 
and other previous investigations are not highly reddened, 
and the careful distance determination by \citet{Genovali-2014} gives
an error size of 5~\% or smaller, i.e.~0.1~mag in distance modulus. 
This error corresponds to roughly $\Delta \RGC \sim 0.2$~kpc at several kpc,
so that the effect on the metallicity gradient is very small.  
For Cepheids affected by severe interstellar extinction, 
like those in \citet{Tanioka-2017}, large errors due to the extinction correction,
{$\sim$}0.5~mag or even larger, would introduce large errors in $\RGC$
(1~kpc at the distance of 5~kpc and more than 2~kpc at $D=10$~kpc).
Such large errors would limit the discussions on the chemical structure of
the disk traced by Cepheids, and efforts to reduce the uncertainty
is required.

It is worthwhile to discuss the metallicity effect on the PL relations
of Cepheids in the context of the metallicity gradient.
Recent studies made use of
near-IR PL relations (or period-Wesenheit relations) to
estimate distances to individual Cepheids in a wide range of the disk
\citep[e.g.][]{Genovali-2014}, and the same approach will be adopted
even in the {\it Gaia} era because of the high extinction in the optical
for relatively distant Cepheids. If the metallicity effect on the PL relations
introduces a systematic error into distances to Cepheids depending 
their metallicities, such an error would introduce an additional scatter to
the radial distribution of the Cepheids' metallicities and/or
skew the gradient from the inner, metal-rich, to the outer, metal-poor,
parts of the disk. However, the metallicity effect on the near-IR PL relations
is considered to be negligible \citep[see e.g.][]{Bono-2010,Subramanian-2017}.
Even a very conservative upper limit, 0.1~mag~dex$^{-1}$, for
the near-IR relations would lead to the small relative error of
{$\sim$}0.1~mag, which is negligible as discussed above,
when the most metal-rich ([Fe/H]$\sim +0.5$~dex) 
and the most metal-poor ([Fe/H]$\sim -0.5$~dex) Cepheids are compared.
The study of the metallicity gradient is therefore not significantly
affected by the possible metallicity effect on the PL relations.
Instead, the metallicity gradient provides us with a sample 
of Cepheids with a sufficiently wide range of metallicity  
within a few kpc for which {\it Gaia} will deliver accurate parallaxes.
Such a sample would be useful to provide constraints on
the metallicity effect on the PL relations.

In addition to the more robust estimation of distance,
the availability of new large catalogs of Cepheids in the Galaxy
calls for a new approach in the determination of their chemical abundances. 
In fact, a comprehensive and detailed survey of element
abundances for hundreds and thousands of Cepheids and young open clusters
throughout the disk would offer the
unique opportunity to map how chemically homogeneous
the Galactic ISM is at a given time in a given
place (e.g. spiral arm). This can be achieved with
large (single-object or multi-object) spectroscopic surveys. 
However, such surveys are usually designed to be single-epoch 
in order to maximize the survey coverage, while
Cepheids are pulsating stars affected by dynamical phenomena
(pulsation, shock, etc.), and their spectra change as function of phase.
Thus, we need to correctly quantify 
the spectral variations due to the observed pulsation phase.
This can be achieved by obtaining phase-resolved spectroscopy of well-understood
Cepheids of different periods, which allows for empirical calibration of 
their variations in effective temperature, gravity and radial velocity.
Recent works
\citep{Luck-2011} show that the metallicities of
a given well-known star based on spectra at different epochs show
a reasonable agreement with each other, but small scatters of
about 0.1~dex may still be attributed to a systematic phase-dependent effect
(see, e.g., the latest work in \citealt{Vasilyev-2018}).

It is worthwhile to spend a few more words on the spectroscopic approach 
to study Cepheids.
A new spin in modelling dynamical atmospheres
is required to properly analyze Cepheid spectra. Even when
considering simple LTE (Local Thermodynamic Equilibrium)
1D atmospheric models of giant stars,
the parameter space (low gravity, cool temperature)
where the Cepheids are located, is significantly under-sampled. Thus,
extrapolation algorithms capable of filling the parameter-space model grid
are necessary to derive Cepheid abundances. 
An alternative approach is to use data-driven spectral models,
which can be constructed on the basis of the already available data-sets and
applied blindly to new data-sets \citep{Ness-2015}. 
Given their importance, Cepheids have already been included as 
a specific class of targets in one of the future largest spectroscopic surveys, 
the 4MOST survey (PI: C.~Chiappini), 
which will deliver final abundances for thousands of new Cepheids
(including those which will be discovered by {\it Gaia}).

\section{Concluding Remarks}
\label{sec:Remarks}

Young and intermediate-age distance indicators are crucial to
study the Galactic disk. Among a few kinds of promising
distance indicators, we have mainly discussed Cepheids and open clusters
(and contact binaries therein). 
Cepheids are well established distance indicators;
they have been, for example, the most successful tracers used for
delineating the metallicity gradient of the disk.
Open clusters and contact binaries are also very useful tracers
and their advantage is
that their ages are spread in a wide range (a few Myr to several Gyr)
in contrast to Cepheids being 10--200~Myr old. Making use of
them properly requires detailed studies on individual clusters,
but they would provide
us with important clues into the evolution of the Galactic disk.

When we use distance tracers discovered in large-scale surveys 
to investigate a large volume of the Galactic disk,
the interstellar extinction poses a very challenging problem.
Our review has summarized massive efforts
to characterize the interstellar extinction 
dedicated for a long time by many astronomers,
but it is also clear that many uncertainties remain.
For the rest of this review, we'll discuss two promising
approaches to determine the extinction law with
newly available data sets today and in the future: 
(1) spectroscopic data, and (2) {\it a priori} distances of reddened 
distance indicators.

With the advent of the APOGEE survey, for example,
much progress is going to be made in the studies of the IR extinction law. 
APOGEE is a high resolution \citep[$R \approx 22,500$;][]{Majewski-2017}, 
$H$-band ($1.51~{\mu{\rm m}} < \lambda < 1.70~{\mu{\rm m}}$)
spectroscopic survey of stars in our Galaxy. Using APOGEE spectra,  
we can measure stellar parameters, such as the effective temperature $\Teff$, 
surface gravity $\log g$, and metal abundance $Z$. 
This allows us to precisely determine $\Teff$ 
and the intrinsic colors of late-type giants.
It should be noted that spectroscopic data for early-type stars
has been often used for studying the extinction law
\citep[e.g.][]{Fitzpatrick-2007}. This kind of approach remains important,
while the advantage of using late-type giants is they are ubiquitous
and much more abundant across the disk so that we can study
the spatial variation of the extinction law, if any, almost continuously.
To extend such studies to more obscured regions across the Galactic disk,
new-generation near-IR high-resolution spectrographs,
e.g.\  CRIRES$^+$ \citep{Dorn-2016} and WINERED \citep{Ikeda-2016},
will be useful thanks to their high sensitivity.

\begin{figure}
  \begin{center}
  \includegraphics[width=0.95\textwidth]{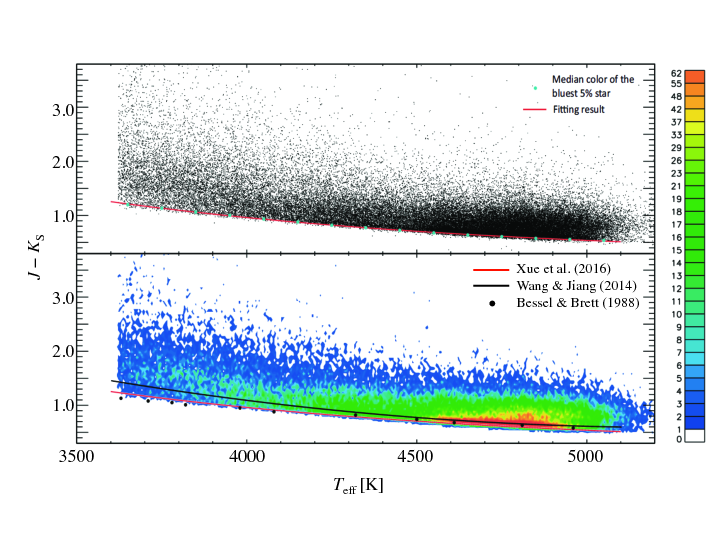}
\caption{
  $J-\Ks$ versus $\Teff$ diagram for 63,330 APOGEE stars.
  The intrinsic, non-reddened colors are derived by using bluest stars
  in this diagram. 
  (Top) The median colors of the bluest 5\,\% of stars are calculated
  in bins of $\Delta \Teff = 100\,$K (cyan dots),
  and they are fit with a quadratic function (red lines).
  (Bottom) The determined intrinsic $J - K_{\mathrm{S}}$ colors 
  by \citet[][red line]{Xue-2016}
  are compared with those from \citet[][black dots]{Bessell-1988} 
  and \citet[][black solid line]{Wang-2014a}. The color scale
  gives the number density (number, see the scale in the legend,
  per 5~K in $\Teff$ per 0.05~mag in color as indicated in the legend).
  Adapted from \citet[][Fig.~1]{Xue-2016}.}
\label{fig:16XueFig1} 
\end{center}
\end{figure}

\begin{figure}
  \includegraphics[width=\textwidth]{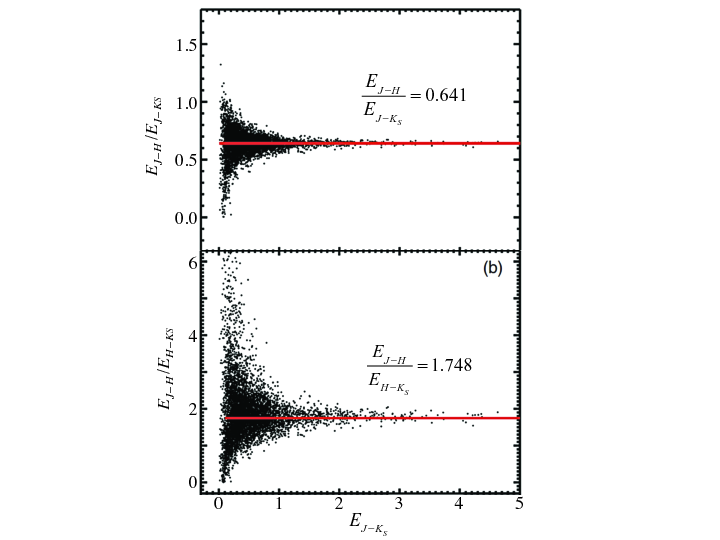}
\caption{
  Distribution of $E_{J-H}/E_{J-K_{\rm s}}$ (top) and $E_{J-H}/E_{H-K_{\rm s}}$ (bottom)
as a function of $E_{J-K_{\rm s}}$ \citep{Wang-2014a}.
The red lines highlight linear fitting results,
$E_{J-H}/E_{J-K_{\rm s}} = 0.64$ (top) and
$E_{J-H}/E_{J-K_{\rm s}} = 1.748$ (bottom). 
There is no apparent tendency in either color-excess ratios
with increasing reddening.
The large dispersion at smaller $E_{J-K_{\rm s}}$ is fully explained
by observational uncertainties.
Adapted from \citet[][Fig.~3]{Wang-2014a}.}
\label{fig:14WangFig4}
\end{figure}

By combining the stellar parameters from APOGEE
and the observed colors from 2MASS,
\citet{Wang-2014a} determined the intrinsic colors of K giants.
The bluest edge of the color index in the $\Teff$-color diagram
is considered to represent the intrinsic colors of the red giants,
i.e., the bluest stars suffer from negligible interstellar extinction
(Fig.~\ref{fig:16XueFig1}).
The color excesses $E_{J-H}$, $E_{H-\Ks}$, and $E_{J-\Ks}$ are
derived for 5942 stars
at $0^{\circ} < l < 220^{\circ}$ and $-5^{\circ} \leq b \leq 5^{\circ}$.
Their interesting results are shown in the color-excess ratio versus
color-excess plot, such as the $E_{J-H}/E_{J-\Ks}$ versus
$E_{J-\Ks}$ diagram (Fig.~\ref{fig:14WangFig4}).
\citet{Wang-2014a} found no tendency of $E_{J-H}/E_{J-\Ks}$
with $E_{J-\Ks}$, for $0 < E_{J-\Ks} < 5$~mag (which corresponds to
$\AKs$ less than 2.5~mag).
A systematic change is not found in 
$E_{J-H}/E_{J-\Ks}$ and $E_{J-H}/E_{H-\Ks}$,
which are often used as a measure of the near-IR extinction law.
The dispersion at low $E_{J-\Ks}$ is likely explained fully 
by observational errors.
\citet{Wang-2014a} claimed that these results confirm
the ``universality'' of the near-IR extinction law in
the $J$, $H$, and $K_{\rm s}$ bands,
from the diffuse to the dense interstellar medium.
The variations found in the previous studies could be due partly to 
differences in the effective wavelength $\lambda_{\mathrm{eff}}$
in different photometric systems, 
or to effects of stellar metallicity,
which was restricted to ${\rm [Fe/H]} > -1.0$
in \citet{Wang-2014a}.
\citet{Schultheis-2015} also found no evidence for any trend of 
the variation of the near-IR extinction law.
They used stellar properties derived from the Gaia-ESO survey,
mainly to probe the 3D distribution of the interstellar extinction
in our Galaxy. The fields of the survey are located at high Galactic latitude,
$|b| > 20^{\circ}$.
They show a plot of $E_{J-H}/E_{H-\Ks}$ versus angle from the GC,
and claim no variation of the extinction coefficient
along different lines of sight.

Such a spectroscopic approach has been extended to the mid-IR wavelengths.
\citet{Xue-2016} applied the method of \citet{Wang-2014a},
i.e.~estimating the intrinsic stellar colors in a spectroscopic way 
for studying the extinction law, to mid-IR data.
The mid-IR data sets they used are from
{\it Spitzer}/IRAC, {\it Spitzer}/MIPS, {\it WISE}, and {\it AKARI}.
In their analysis, G- and K-type giants are used.
To examine the variation of the extinction law,
\citet{Xue-2016} divided their data sets into three groups:
$\AKs < 0.5$, $0.5 \leq \AKs < 1.0$, and $\AKs \geq 1.0$.
Then they compared the color-excess ratios $E_{\Ks-\lambda}/E_{J-\Ks}$
in the three groups, for the {\it WISE}, {\it AKARI}, and {\it Spitzer} nine bands.
They also compared the extinction laws in the form $A_{\lambda}/\AKs$
under the assumption that 
$A_J/\AKs = 2.72$ is universal for the three groups.
\citet{Xue-2016} found remarkably constant color-excess ratios
among the three groups. This is also seen in $A_{\lambda}/\AKs$.

\citet{McClure-2009}, in contrast, detected variations in
mid-IR extinction using {\it Spitzer}/MIPS spectra of  
31 late-type giants behind dense molecular clouds.
Their conclusion is that the median curves for specific ranges of
the amount of extinction characterize the shape of the extinction law.
They claimed that they found a ``real'' variation 
of the extinction law,
as a function of $A_K$ in the mid-IR waverange;
the extinction law also becomes flat with increasing $A_K$.
The difference from the results in
\citet{Xue-2016} may simply reflect the difference between
the extinction by relatively diffuse interstellar medium and
that by dense molecular clouds.
The targets of \citet{McClure-2009} are molecular clouds
in star forming regions,
and the clouds are dense enough for grain growth to occur.
As predicted by grain models \citep[e.g.][]{Weingartner-2001},
the grain growth is likely to alter the shape of the extinction law.
On the other hand, \citet{Wang-2014a} and \citet{Xue-2016} do not
distinguish among interstellar environments.
Even though there still remain uncertainties,
massive spectroscopic data available today and in the future
may be critical to determine the extinction law in a precise and robust way.

Besides the spectroscopic approaches, reddened objects with
known distance are also useful to determine the extinction law.
In order to give observational constraints on
the total-to-selective extinction ratio (e.g.~$\AKEHK$),
distance indicators whose distances can be addressed 
by more than one methods are very useful.
For example, \citet{Matsunaga-2016,Matsunaga-2017} adopted
the extinction law in \citet{Nishiyama-2006}, and rejected some other laws,
based on observational data for Cepheids in the Nuclear Stellar Disk,
within {$\sim$}230~pc around the GC \citep{Launhardt-2002}.
For these Cepheids, while distance estimation based on
the PL relation of Cepheids is affected by the extinction law,
their distances should agree with that of GC within the size of
the Nuclear Stellar Disk because of their membership to this
relatively small-sized disk.
There are many independent estimates of the distance to the GC,
as discussed in \citet[$8.3\pm 0.4$~kpc]{deGrijs-2016}, including 
Keplerian stellar orbits around the central supermassive black hole
\citep{Boehle-2016,Gillessen-2017,Parsa-2017} which is unaffected
by the extinction law. Considering how different extinction laws
($\AKEHK$ to be more specific) lead to different estimates of distance
in comparison with the assumed distance from that of GC limits
an acceptable range of the extinction law.  
\citet{Matsunaga-2016} concluded that 
$\AKEHK=1.44$ found by \citet{Nishiyama-2006} 
or slightly smaller values would be consistent with
their observational data for the Cepheids in the Nuclear Stellar Disk.
Larger $\AKEHK$ values would put the four Cepheids (with $\EHKs\sim 1.6$~mag)
significantly more distant than expected; $\AKEHK=1.61$ in
\citet{Nishiyama-2009} lead to $\AKs$ larger by
$1.7 \times (1.61-1.44) \sim 0.3$~mag, 15~\% larger in distance,
or even a larger difference is introduced by $\AKEHK=1.83$
in \citet{Cardelli-1989}.

\begin{figure}
  \begin{center}
    \includegraphics[width=\textwidth]{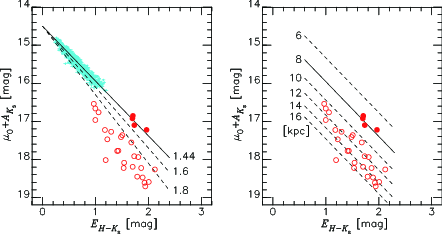}
    \caption{
      (Left) Reddening vectors on the color-magnitude diagram
      with different coefficients of $\AKEHK$
      starting from the distance of the Galactic center, 8~kpc, at
      the zero color excess, $\EHKs=0$~mag. The sequence of cyan points
      near the top-left corner, taken from \citet{Nishiyama-2006},
      indicates the RC peaks affected by various amounts of reddening.
      Filled circles indicate
      the four classical Cepheids in the nuclear stellar disk
      \citep{Matsunaga-2011,Matsunaga-2015}, and open circles indicate
      the other classical Cepheids reported in \citet{Matsunaga-2016}.
      (Right)~Same as the left panel, but lines have the slope of
      $\AKEHK=1.44$ and correspond to different distances from 6 to 16~kpc.
      Adapted from \citet[][Fig.~6]{Matsunaga-2017}.}
    \label{fig:Matsunaga17Fig6a}
  \end{center}
\end{figure}

\citet{Nishiyama-2006} themselves obtained $\AKEHK$ and other
parameters of the $\JHKs$ extinction law based on a similar but slightly
different principle. They compared RC peaks toward many sightlines
for which the amounts of extinction are spread in a wide range,
0.3--1.2 in $\EHKs$. They assumed that the RC peaks along
the sightlines toward the bulge trace the same distance,
because of the concentration toward the center of the bulge,
even with the different amounts of extinction.
The difference of $\Ks$ magnitudes {$\sim$}1.3 between the least and
most reddened RC peaks observed by \citet{Nishiyama-2006} leads to
their extinction law, $\AKEHK = 1.44 \simeq 1.3/(1.2-0.3)$.
Note that they didn't use the distance to the GC (or the bulge) but
used the relative extinction and reddening of RC peaks assumed to
be at the same distance. The results of \citet{Matsunaga-2016} and
\citet{Nishiyama-2006} are summarized in an illustrative way in
Fig.~\ref{fig:Matsunaga17Fig6a}.
These works have demonstrated that 
distance indicators play a crucial role in determining the extinction law
especially when their distances can be estimated by more than one method;
{\it Gaia} parallaxes in the near future, for example, will provide us with
great data sets also for problems of the extinction law.

\begin{acknowledgements}
We thank the organizers and participants of ISSI-BJ for
the inspiring discussions therein.
NM is grateful to Grant-in-Aid, KAKAENHI No.~26287028, from
the Japan Society for the Promotion of Science (JSPS).
RdG acknowledges research support from the National Natural Science Foundation
of China (grants U1631102, 11373010, and 11633005).
SN was supported by JSPS KAKENHI (Nos.~25707012, 15K13463). 
\end{acknowledgements}

\bibliographystyle{spbasic}      


\end{document}